\documentclass[fleqn,usenatbib]{mnras}

\usepackage{newtxtext,newtxmath}

\usepackage[T1]{fontenc}

\DeclareRobustCommand{\VAN}[3]{#2}
\let\VANthebibliography\thebibliography
\def\thebibliography{\DeclareRobustCommand{\VAN}[3]{##3}\VANthebibliography}

\usepackage{graphicx}	
\usepackage{amsmath}	

\DeclareMathOperator\sign{sgn}

\title[neutron star - black hole kilonovae]{A multi-messenger model for neutron star - black hole mergers}

\author[B. P. Gompertz et al.]{
B. P. Gompertz,$^{1,2}$\thanks{E-mail: b.gompertz@bham.ac.uk (BPG)}
M. Nicholl,$^{1,2,3}$
J. C. Smith,$^{1}$
S. Harisankar,$^{1}$
G. Pratten,$^{1,2}$
P. Schmidt,$^{1,2}$
and
\newauthor G. P. Smith$^{1}$
\\
$^{1}$School of Physics and Astronomy, University of Birmingham, Birmingham, B15 2TT, UK\\
$^{2}$Institute for Gravitational Wave Astronomy, University of Birmingham, Birmingham, B15 2TT, UK\\
$^{3}$Astrophysics Research Centre, School of Mathematics and Physics, Queen’s University Belfast, Belfast BT7 1NN, UK\\
}

\date{Accepted XXX. Received YYY; in original form ZZZ}

\pubyear{2022}

\begin{document}
\label{firstpage}
\pagerange{\pageref{firstpage}--\pageref{lastpage}}
\maketitle

\begin{abstract}
We present a semi-analytic model for predicting kilonova light curves from the mergers of neutron stars with black holes (NSBH). The model is integrated into the {\sc mosfit} platform, and can generate light curves from input binary properties and nuclear equation-of-state considerations, or incorporate measurements from gravitational wave (GW) detectors to perform multi-messenger parameter estimation. The rapid framework enables the generation of NSBH kilonova distributions from binary populations, light curve predictions from GW data, and statistically meaningful comparisons with an equivalent BNS model in {\sc mosfit}. We investigate a sample of kilonova candidates associated with cosmological short gamma-ray bursts, and demonstrate that they are broadly consistent with being driven by NSBH systems, though most have limited data. We also perform fits to the very well sampled GW170817, and show that the inability of an NSBH merger to produce lanthanide-poor ejecta results in a significant underestimate of the early ($\lesssim 2$ days) optical emission. Our model indicates that NSBH-driven kilonovae may peak up to a week after merger at optical wavelengths for some observer angles. This demonstrates the need for early coverage of emergent kilonovae in cases where the GW signal is either ambiguous or absent; they likely cannot be distinguished from BNS mergers by the light curves alone from $\sim 2$ days after the merger. We also discuss the detectability of our model kilonovae with the Vera C. Rubin Observatory's Legacy Survey of Space and Time (LSST).
\end{abstract}




\section{Introduction}

Our understanding of compact object mergers has made significant advances following the advent of gravitational-wave (GW) astronomy, including the first ever detection in GW of a binary black hole (BBH) merger \citep{Abbott16}, binary neutron star (BNS) merger \citep{Abbott17_GW} and most recently the merger of a neutron star -- black hole (NSBH) system \citep{Abbott21_NSBH}. Where neutron stars (NS) are involved, accompanying electromagnetic (EM) signals like short gamma-ray bursts \citep[SGRBs; e.g.][]{Paczynski86,Kouveliotou93,Berger14} and kilonovae \citep{Li98,Rosswog05,Metzger10,Barnes13,Metzger19} are expected. Both became confirmed counterparts of BNS mergers with the coincident detections of GW170817 \citep{Abbott17_GW}, GRB 170817A \citep{Goldstein17,Hallinan17,Margutti17,Savchenko17,Troja17,DAvanzo18,Lyman18,Margutti18,Mooley18,Troja18b,Lamb19b} and the kilonova AT2017gfo \citep{Andreoni2017,Arcavi2017,Chornock17,Coulter17,Cowperthwaite17,Drout17,Evans17,Kasliwal17b,Lipunov2017,McCully2017,Nicholl17,Pian17,Shappee2017,Smartt17,Soares-Santos17,Tanvir17,Utsumi2017,Valenti2017,Villar17}.

The association of kilonovae with BNS mergers has important implications for the production of heavy elements in the Universe. These thermal transients are powered by the radioactive decay of unstable heavy elements assembled by rapid neutron capture ($r$-process) nucleosynthesis following the merger \citep{Lattimer74,Eichler89,Freiburghaus99}. Modelling of the GW170817 kilonova indicates that BNS mergers may be the dominant source of $r$-process elements in the Universe \citep{Rosswog18}. However, comparisons with kilonova candidates associated with cosmological SGRBs \citep{Berger13,Tanvir13,Yang15,Jin15,Jin16,Kasliwal17,Jin18,Troja18,Eyles19,Lamb19,Troja19,Jin20,Fong21,O'Connor21,Rastinejad22,Troja22,Levan23} imply that the yield of $r$-process elements is highly variable between events \citep{Gompertz18,Ascenzi19,Rastinejad21}. In addition, significant uncertainties remain in the measured BNS merger rate. Estimates from GW events \citep[$320^{+490}_{-240}$\,Gpc$^{-3}$\,yr$^{-1}$;][]{Abbott21} are hampered by the low number of detections to date, while inferences from the rate of short GRB detections \citep[$270^{+1580}_{-180}$\,Gpc$^{-3}$\,yr$^{-1}$;][]{Fong15} must account for the jet opening angle distribution, which is poorly constrained. The exact contribution BNS mergers make to the $r$-process census is therefore highly uncertain.

A growing number of studies seek to minimise this uncertainty through simultaneous modelling of both the EM and GW observations, where available \citep{Margalit17,Margalit19,Barbieri19,Coughlin19,Dietrich20,Breschi21,Nicholl21,Raaijmakers21}. Measurements of the binary and post-merger remnant from GW interferometers like advanced LIGO \citep{LSC15}, advanced Virgo \citep{Acernese15} and KAGRA \citep{KAGRA19} can be combined with observations of the subsequent transient from EM observatories and synthesised into tighter posterior distributions for parameters that impact the nucleosynthesis yield \citep[e.g][]{Abbott17_multimessenger}. They can also provide more stringent constraints on the NS equation-of-state.

A significant additional uncertainty in the Universal $r$-process census is the contribution made by NSBH mergers \citep[see e.g.][]{Chen21}. Such events are theoretically capable of driving SGRBs and kilonovae \citep[e.g.][]{Rosswog05,Tanaka14,Paschalidis15,Desai19} if the NS is disrupted before plunging into the BH, and some candidate NSBH-driven events have been proposed in the literature \citep[e.g.][]{Troja08,Yang15,Jin16,Kawaguchi16,Gompertz20,Zhu22}. However, the mass of disrupted material that remains outside of the remnant BH event horizon is expected to be low if the mass ratio of the binary is high and/or the magnitude of the orbit-aligned component of the pre-merger BH spin is low or negative \citep{Foucart14,Pannarale14,Kawaguchi16,Foucart18}. The early GW-detected NSBH merger events \citep{Abbott21_NSBH} and candidates \citep{GWTC-2,GWTC-2.1,GWTC-3} exhibit total masses and mass ratios that are suitable for NS disruption. However, the measured BH spins are consistent with zero, and the mergers are not expected to be EM bright \citep{Dichiara21,Fragione21,Mandel21,Zhu21,Gompertz22b}. They appear to derive from the isolated binary evolution channel \citep{Broekgaarden21,Broekgaarden21b}, though potentially via a non-standard pathway \citep{Gompertz22b}. The exception is GW191219\_163120 \citep{GWTC-3}, whose large mass ratio implies that the binary may have formed through dynamical capture \citep{Gompertz22b}.

While zero BH spin at the point of merger is a common prediction from population synthesis modelling \citep[e.g. for BBH systems;][]{Qin18,Fuller19}, pathways to higher spin systems are possible through weak core-envelope coupling in BH progenitor stars, or tidal interactions following BH formation \citep{Steinle21,Steinle23}. Should such systems be realised in nature, they are expected to be accompanied by bright kilonovae with nucleosynthesis yields up to 10x greater per event than that expected from BNS mergers \citep{Tanaka14}. Their still-uncertain merger rate density may be comparable to that of BNS mergers, but could also be significantly lower \citep{Mapelli18,Eldridge19,Belczynski20,Abbott21_NSBH}. The potential contribution of NSBH mergers to Universal $r$-process production therefore ranges from none at all to being the dominant production sites of lanthanides and actinides through cosmic time. Calibrating their influence will require further detections of events in GW during LIGO-Virgo-KAGRA (LVK) observing runs to constrain merger rates, as well as EM detections or stringent limits on emission that translates to meaningful measurements or constraints on $r$-process yields. This is best achieved through GW-EM multi-messenger modelling.

In this paper, we present a semi-analytic forward model for NSBH-driven kilonovae that predicts light curves from the binary configuration and NS equation-of-state. The relative simplicity of our model compared to more simulation-based alternatives means that it is optimised for quickly generating light curves for arbitrary parameters, fitting to data, predicting populations, or marginalising over unconstrained parameters. By providing the model within the {\sc mosfit} framework \citep{Guillochon18}, it is publicly available for easy use and adaptation, and trivial to perform model comparison against an equivalent BNS model \citep{Nicholl21}, e.g. for modelling mass-gap systems, or when no GW data are available. In the absence of GW observations, fitting to kilonova light curves affords constraints on the properties of the progenitor binary. Any available GW information can be included in the priors to enable multi-messenger inference of the merger, and tight constraints on the nucleosynthesis yield and equation-of-state.

Our paper is structured as follows. The model is described in Section~\ref{sec:model} and compared to a well-sampled subset of SGRB kilonovae to see if any are compatible with being NSBHs in Section~\ref{sec:GRB-KNe}. We perform fits to the GW-EM multi-messenger dataset of GW170817 to search for a self-consistent NSBH solution in Section~\ref{sec:170817}. We discuss the implications our model has for the detectability of NSBH kilonovae with the Vera C. Rubin Observatory in Section~\ref{sec:LSST}. Finally, we present our conclusions in Section~\ref{sec:conclusions}. Magnitudes are in the AB system unless otherwise stated.

\section{Model Description}\label{sec:model}

\begin{figure*}
    \centering
    \includegraphics[width=17cm]{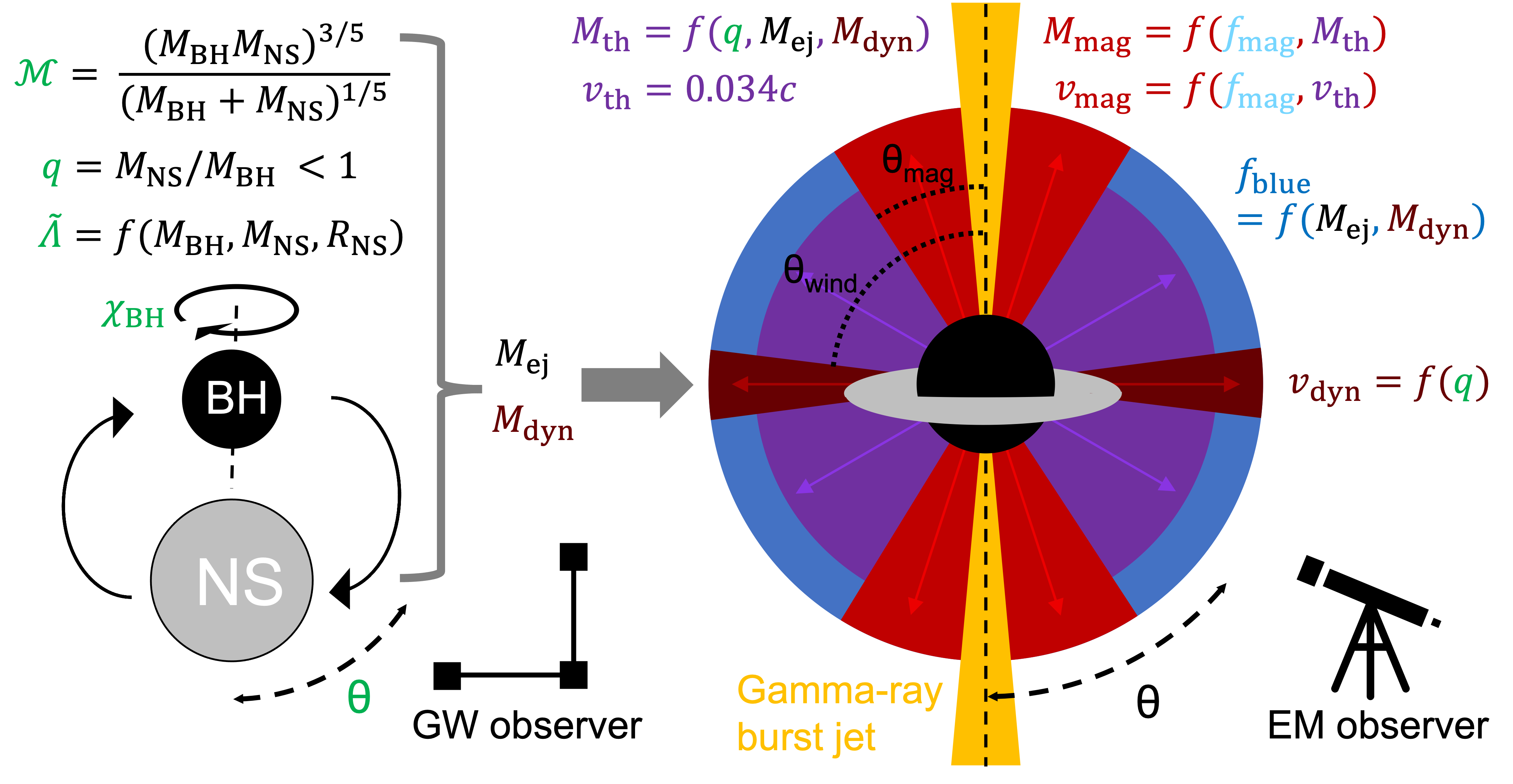}
    \caption{Schematic of the model. The five measured GW parameters are shown in green. The total ejecta mass \citep[$M_{\rm ej}$, Equation~\ref{eq:ejecta};][]{Foucart18} and dynamical ejecta mass \citep[$M_{\rm dyn}$, Equation~\ref{eq:dyn};][]{Kruger20} are functions of the binary properties, and influence the kilonova light curve evolution. The masses and velocities of individual emission components are shown in their respective colours, along with their dependencies. The grey opacities for each component are $\kappa_{\rm blue} = 1$\,cm$^2$g$^{-1}$, $\kappa_{\rm th} = 5$\,cm$^2$g$^{-1}$ and $\kappa_{\rm mag} = \kappa_{\rm dyn} = 10$\,cm$^2$g$^{-1}$.}
    \label{fig:cartoon}
\end{figure*}

A schematic overview of our model is shown in Figure~\ref{fig:cartoon}. For an electromagnetic transient to be produced, the NS must be disrupted by the tidal forces exerted upon it by the BH in the final stages of inspiral, with some mass remaining outside of the BH event horizon. Tidal disruption occurs if the NS overflows its Roche lobe at distances greater than the innermost stable circular orbit (ISCO) of the BH. This radius can be expressed as \citep[cf.][]{Bardeen72}:
\begin{equation}
    \hat{R}_{\rm ISCO} = 3 + Z_2 - \sign(\chi_{\rm BH}) \sqrt{(3 - Z_1)(3 + Z_1 + 2Z_2)},
\end{equation}
where $\hat{R}_{\rm ISCO} = R_{\rm ISCO}/M_{\rm BH}$ is the normalised ISCO radius, $M_{\rm BH}$ is the BH mass, $\chi_{\rm BH}$ is the orbit-aligned component of the BH's dimensionless spin parameter, $Z_1 = 1 + (1 - \chi_{\rm BH}^2)^{1/3} \big[(1 + \chi_{\rm BH})^{1/3} + (1 - \chi_{\rm BH})^{1/3}\big]$ and $Z_2 = \sqrt{3 \chi_{\rm BH}^2 + Z_1^2}$.

An analytical fitting function for the mass of the material that remains outside of the BH event horizon was derived by \citet{Foucart18}. The fitting function was calibrated to 75 numerical relativity simulations \citep[compiled from][]{Etienne09,Foucart11,Kyutoku11,Foucart12,Foucart13,Lovelace13,Foucart14,Kyutoku15,Brege18}, and gives an ejected mass of
\begin{equation}\label{eq:ejecta}
    M_{\rm ej} = M_{\rm NS}^b\bigg[\max\bigg(\alpha \frac{1-2C_{\rm NS}}{\eta^{1/3}} - \beta \hat{R}_{\rm ISCO} \frac{C_{\rm NS}}{\eta} + \gamma , 0\bigg)\bigg]^{\delta},
\end{equation}
where the four fitting parameters were found to be $\alpha = 0.406$, $\beta = 0.139$, $\gamma = 0.255$ and $\delta = 1.761$. Equation~\ref{eq:ejecta} parameterises the ejected mass in terms of $\hat{R}_{\rm ISCO}$, $\eta = (1+1/q)^{-2}q^{-1}$ (where $q = M_{\rm NS}/M_{\rm BH}$ is the binary mass ratio), the compactness of the NS
\begin{equation}
    C_{\rm NS} = GM_{\rm NS} / (R_{\rm NS}c^2),
\end{equation}
and its baryonic mass \citep[cf.][]{Lattimer01}:
\begin{equation}
    M_{\rm NS}^b = M_{\rm NS}\bigg(1 + \frac{0.6C_{\rm NS}}{1-0.5C_{\rm NS}}\bigg).
\end{equation}
The ejected mass of the merger is therefore primarily a function of the orbit-aligned component of the BH's spin, the binary mass ratio, and the NS equation-of-state.

\subsection{Dynamical ejecta}\label{sec:dynamical}

\citet{Kruger20} developed an analytical fitting function for the mass of material ejected dynamically from an NSBH merger:
\begin{equation}\label{eq:dyn}
    \frac{M_{\rm dyn}}{M_{\rm NS}^b} = a_1 q^{-n_1}\frac{1-2C_{\rm NS}}{C_{\rm NS}} - a_2 q^{-n_2} \frac{R_{\rm ISCO}}{M_{\rm BH}} + a_4.
\end{equation}
The best-fitting parameters, validated against simulations by \citet{Kawaguchi15} and \citet{Foucart19}, were found to be $a_1 = 0.007116$, $a_2 = 0.001436$, $a_4 = -0.02762$, $n_1 = 0.8636$, and $n_2 = 1.6840$. The average velocity of this ejecta was found to be an inverse function of q \citep{Kawaguchi16}:
\begin{equation}
    v_{\rm dyn} = (0.01533q^{-1} + 0.1907)c.
\end{equation}

The dynamical ejection of matter is primarily driven by tidal torque, and is therefore typically distributed within $10^{\circ}$ - $20^{\circ}$ of the orbital plane \citep[e.g.][]{Kawaguchi15,Kyutoku15}. For simplicity we assume an axisymmetric distribution \citep[see however][]{Kyutoku15,Kawaguchi16}. The tidal dynamical ejecta experience only weak neutrino irradiation \citep{Kyutoku18}, meaning that the electron fraction is expected to be low \citep[$Y_e \lesssim 0.1$][]{Foucart14,Metzger14,Kyutoku18}. We model the dynamical ejecta with a gray opacity of $\kappa_{\rm dyn} = 10$\,cm$^2$g$^{-1}$ \citep{Tanaka13,Kawaguchi16,Kasen17,Tanaka20}.

\subsection{Disc winds}\label{sec:winds}

\subsubsection{Thermally driven wind}

Combining the work of \citet{Foucart18} and \citet{Kruger20}, we obtain the disc mass:
\begin{equation}\label{eq:discmass}
    M_{\rm disc} = M_{\rm ej} - M_{\rm dyn}.
\end{equation}
Hydrodynamic simulations show that some of the post-merger disc surrounding the remnant BH is driven away in neutron-rich winds by viscous heating and nuclear recombination  \citep[e.g.][]{Fernandez13,Fernandez15,Just15,Fernandez20,Fujibayashi20}. The fraction of the disc that is ejected this way was shown to be a linear function of the disc compactness by \citet{Fernandez20}, and was parameterised as a function of the binary mass ratio by \citet{Raaijmakers21} as
\begin{equation}\label{eq:discfrac}
    \xi = \frac{M_{\rm th}}{M_{\rm disc}} = \xi_1 + \frac{\xi_2 - \xi_1}{1 + e^{1.5(1/q-3)}}.
\end{equation}
We assume $\xi_1 = 0.18$ and $\xi_2 = 0.29$, the median values given in \citet{Raaijmakers21}.

We combine Equations \ref{eq:discmass} and \ref{eq:discfrac} to obtain the mass of material driven from the disc by thermal pressure ($M_{\rm th}$). This material is assumed to have an average velocity of $v_{\rm therm} = 0.034$c \citep{Fernandez20}. However, the outflow velocity is sensitive to the assumed viscosity parameter in the simulation, with higher viscous coefficients associated with more efficient acceleration of matter in the outer accretion disc \citep[e.g.][]{Fujibayashi20}.

The electron fraction of the thermal wind is typically found to be in the range $0.25 \leq Y_e \leq 0.35$ \citep[e.g.][]{Foucart15,Fernandez20,Fujibayashi20}. This corresponds to a gray opacity of $\kappa \lesssim 5$\,cm$^2$g$^{-1}$ in \citet{Tanaka20}, with the true value tending towards the lower end of this range when temperatures are below 5000K \citep[e.g. $\kappa = 1$\,cm$^2$g$^{-1}$ in][]{Kasen17}. \citet{Fernandez20} find that a significant portion of this wind has a lanthanide and actinide mass fraction $X_{(La+Ac)} < 10^{-4}$. Motivated by this, we model the thermal wind as a two component mixture model featuring a leading blue edge with $\kappa = 1$\,cm$^2$g$^{-1}$ enveloping a redder core with $\kappa = 5$\,cm$^2$g$^{-1}$. The blue mass fraction ($f_{\rm blue}$) was found to monotonically increase with disc mass by \citet{Fernandez20}, so we calculate $f_{\rm blue}$ from $M_{\rm disc}$ (Equation~\ref{eq:discmass}) using a first-order polynomial fit to the data in their Table~2, noting the large scatter induced by varying the BH mass:
\begin{equation}
    f_{\rm blue} = 0.20199 \log_{\rm 10}(M_{\rm disc}) + 1.12692.
\end{equation}

\subsubsection{Magnetically driven wind}

The inclusion of magnetic fields in three-dimensional general-relativistic magnetohydrodynamic (GRMHD) models \citep{Siegel17,Siegel18,Christie19,Fernandez19} has revealed a second outflow in the form of an MHD-mediated wind. This results in twice as much ejecta mass, a higher average ejecta velocity and a lower average electron fraction ($Y_e$) when compared to equivalent hydrodynamic simulations \citep{Fernandez19}. The mass ejected by magnetic processes depends on the geometry of the post-merger magnetic field \citep{Christie19}. More poloidal configurations eject more mass, and with higher velocities \citep[cf.][]{Fernandez19}, while preferentially toroidal fields generate very little magnetically driven ejecta \citep[cf.][]{Siegel18}. \citet{Fernandez19} find that the magnetically-driven outflow has a velocity $v > 0.1c$, in excess of the maximum velocity seen in hydrodynamic simulations.

We include this second wind component in our fiducial model with the ignorance parameter $f_{\rm mag}$, which accounts for the unknown magnetic field configuration. A fully poloidal field has $f_{\rm mag} = 1$, while lower values represent more toroidal field geometries. It is applied as a fraction of the thermal wind ejecta mass derived in Equation~\ref{eq:discfrac}: $M_{\rm mag} = f_{\rm mag} M_{\rm th}$, and as $v_{\rm mag} = f_{\rm mag} 0.22c$, where $0.22c$ is the average velocity of the faster bimodal component in the fully poloidal field geometry of \citet{Fernandez19}. The velocity floor is set equal to the thermal wind velocity ($v_{\rm mag} \geq 0.034c$). The magnetic wind component has $Y_e \sim 0.1$, corresponding to $\kappa = 10$\,cm$^2$g$^{-1}$. This low electron fraction is maintained because the magnetic wind is driven from the disc towards the poles before it is significantly impacted by neutrino irradiation.

The inclusion of the magnetically driven wind means that our model predicts lanthanide rich and therefore optically faint emission. The assumption of high opacity and suppressed optical emission is common in semi-analytical models for NSBH kilonovae \citep[e.g.][]{Barbieri19,Raaijmakers21}, where the lack of neutrino irradiation from a remnant NS means that the electron fraction is expected to be low. However, assuming neutrino irradiation from the inner accretion disk is sufficient to significantly raise the electron fraction can have a marked effect on the early optical light curve \citep[e.g.][]{Zhu20}.

\subsection{Conversion to light curves}

Our model calculates $r$-process ejecta masses and velocities from the input binary configuration. In order to convert them to kilonova light curves, we incorporate the NSBH ejecta model as a package in MOSFiT \citep{Guillochon18}. $r$-process masses and velocities are converted to light curves through pre-existing MOSFiT modules, including semi-analytical models for heating rates and deposition \citep{Korobkin12,Barnes16,Cowperthwaite17,Villar17,Metzger19}, an approximation of photon diffusion through the ejecta \citep{Arnett82}, and self-consistent evolution of the photospheric radius \citep{Nicholl17b}.

The process to generate light curves is as follows \citep[c.f.][]{Villar17}: for each ejecta component, the radioactive heating rate with time is approximated \citep{Korobkin12} as:
\begin{equation}
    L_{\rm in} = 4 \times 10^{18}M_{\rm r} \times \bigg[ 0.5 - \pi^{-1} \arctan \bigg(\frac{t-1.3}{0.11}\bigg) \bigg]^{1.3} \mbox{erg\,s$^{-1}$,}
\end{equation}
where $M_{\rm r}$ is the mass of the $r$-process ejecta. This neglects any contribution from fallback accretion onto the remnant, which is expected to be prevented by winds from the disk \citep[e.g.][]{Fernandez13}. Not all of this energy is available to power the kilonova because only a fraction $\epsilon_{\rm th}$ thermalises within the plasma. As the ejecta become more diffuse with time, the efficiency of thermalisation decreases. This effect is approximated analytically \citep{Barnes16} as
\begin{equation}
    \epsilon_{\rm th}(t) = 0.36 \bigg[ e^{-at} + \frac{\ln(1 + 2bt^d)}{2bt^d}\bigg],
\end{equation}
where $a$, $b$ and $d$ are constants that depend on the mass and velocity of the ejecta and are obtained by interpolating Table 1 of \citet{Barnes16}.

Homologous expansion of the ejecta and central energy deposition are assumed, so that the observed bolometric luminosity of each ejecta component can be calculated as \citep{Arnett82}
\begin{equation}
    L_{\rm bol}(t) = \exp \bigg( \frac{-t^2}{t_d^2} \bigg) \times \int^t_0 L_{\rm in}(t)\epsilon_{\rm th}(t)\exp(t^2/t_d^2)\frac{t}{t_d}dt,
\end{equation}
where $t_{\rm d} \equiv \sqrt{2\kappa M_{\rm r}/\beta vc}$ and $\beta = 13.4$ is a dimensionless constant. The spectral energy distribution (SED) of each component is calculated by assuming blackbody radiation with luminosity $L_{\rm bol}$ and a photospheric radius determined using the prescription in \citet{Nicholl17b}. The SEDs of individual emitting components are then summed in a ratio determined by their relative areas subtended to the observer (see Section~\ref{sec:geometry}), and transformed into light curves in individual photometric filters using the transmission curves available on the Spanish Virtual Observatory (SVO) filter profile service\footnote{\href{http://svo2.cab.inta-csic.es/theory/fps/}{http://svo2.cab.inta-csic.es/theory/fps/}}.

The assumptions built into the light curve creation add further systematic uncertainties to the model. First, the use of a gray opacity is a simplified approximation of the complex electron orbital transitions that are present in heavy element like lanthanides and actinides. A full treatment of the (still incomplete) available atomic data \citep[e.g.][]{Smartt17,Watson19,Gillanders22} may produce different evolution, especially at late times when the assumption of local thermal equilibrium breaks down \citep[e.g.][]{Gillanders23a,Gillanders23b,Hotokezaka23,Levan23}. Detailed nuclear heating with density-dependent thermalisation has also been shown to introduce variability in the bolometric luminosity of kilonovae \citep{Korobkin21,Bulla23}, which may impact parameter estimation from fitting. Finally, more detailed treatment of radiative transfer \citep[e.g.][]{Bulla19} will impact the light curve when compared to the simplified treatment of photon diffusion employed here.

\subsection{Geometry}\label{sec:geometry}

The outflow geometry is structured in a similar fashion to \citet{Nicholl21}. We assume an axially symmetric kilonova and model each emission component as a cutout with a conical polar cap defined in terms of its half-opening angle $\theta_{\rm open}$. Emitting regions are constructed following the formalism of \citet{Darbha20}, where the luminosity of each region is scaled to the area of the caps projected to an observer at a viewing angle $\theta_{\rm obs}$. See their Appendix A for the mathematical expressions. For fiducial parameters we assume that the magnetic wind is restricted to polar regions with $\theta_{\rm mag} = 45^{\circ}$, the thermal wind occupies moderate latitudes ($\theta_{\rm wind} = 80^{\circ}$) and the dynamical ejecta sits $\pm 10^{\circ}$ from the equator. A schematic of the model is shown in Figure~\ref{fig:cartoon}. We note that the evolution of the projected area of the photosphere is complicated in the presence of multiple emission components, especially when they are likely non-spherical \citep[e.g.][]{Zhu20, Just22}.

For simplicity, our model assumes that the emitting regions do not interact. This is a reasonable assumption for the tidally ejected dynamical component, but interactions between the thermal and magnetic winds are likely to produce turbulence along their contact interface. However, our assumption is a reasonable approximation for the majority of viewing angles, and the 50:50 contribution of the two emitting regions when viewed along the boundary between them is also likely a reasonable proxy for a mixed emission component. We do not account for the possibility of polar cavities carved out by a relativistic jet launched by the merger, which may expose hot, low opacity material \citep{Nativi21,Klion21}.

One caveat to our model is that it is based on simulations where the BH spin axis and binary orbital axis are aligned. It has been shown that only considering the aligned spin cases still results in accurate estimates of the mass that remains outside of the BH \citep{Foucart13,Kawaguchi15}. However, misalignment may induce spin precession, which breaks symmetry and is likely to result in asymmetric structure in the ejecta \citep[e.g.][]{Kawaguchi15}. This is not captured in our model.

\begin{table}
    \centering
    \begin{tabular}{cccc}
    \hline\hline
    Parameter & Fiducial value & Astrophysical prior & GW170817 prior \\
    \hline
    $\mathcal{M}^a$ ($M_{\odot}$) & 2.22* & [1.0, 6.0] & $1.188^{+0.004}_{-0.002}$ \\
    $q^b$ & 0.28* & [0.1, 1.0] & [0.4, 1.0] \\
    $\Tilde{\Lambda}^c$ & 11.0* & [0.0, 100.0] & [0.0, 700.0] \\
    $\chi_{\rm BH}^d$ & 0.8 & [-1.0, 1.0] & [-0.01, 0.17] \\
    $\cos \theta^e$ & 0.707 & [0.0, 1.0] & [0.883, 1.0] \\
    $\cos \theta_{\rm mag}^f$ & 0.707 & [0.5, 1.0] & [0.5, 1.0] \\
    $\cos \theta_{\rm wind}^g$ & 0.174 & [0.0, 0.342] & [0.0, 0.342] \\
    $f_{\rm mag}^h$ & 1.0 & [0.1, 1.0]& [0.1, 1.0] \\
    $\log N_H^i$ & 19.0 & [19.0, 23.0] & [19.0, 23.0] \\
    \hline\hline
    Parameter & Fiducial value \\
    \hline
    $M_{\rm dyn}$ & $0.047$\,M$_{\odot}$ \\
    $M_{\rm mag}$ & $0.036$\,M$_{\odot}$ \\
    $M_{\rm th}$ & $0.036$\,M$_{\odot}$ \\
    $v_{\rm dyn}$ & $0.25$c \\
    $v_{\rm mag}$ & $0.22$c \\
    $v_{\rm th}$ & $0.034$c \\
    \hline\hline
    \end{tabular}
    \caption{\textbf{Upper:} The free parameters of the NSBH kilonova model, their assumed fiducial values, and the prior ranges used when fitting. Bracketed values indicate a flat prior distribution, while Gaussian priors are given as median values with one sigma confidence intervals. The GW170817 prior set uses the high spin priors from \citet{Abbott17_GW}.\newline \textbf{Lower:} The masses and velocities produced by the fiducial model for each ejecta component (see Figure~\ref{fig:cartoon}). \newline $^a$Chirp mass. $^b$Mass ratio. $^c$Effective tidal deformability of the binary. $^d$Orbit-aligned component of the BH spin. $^e$Observer viewing angle. $^f$Opening angle of the magnetic wind. $^g$Opening angle of the wind -- dynamical ejecta boundary. $^h$Magnetic wind fraction. $^i$Hydrogen column density in host galaxy (proportional to extinction). *These parameters describe a binary with a $5M_{\odot}$ BH and a $1.4M_{\odot}$ NS with a 12\,km radius.}
    \label{tab:model}
\end{table}

\begin{figure*}
    \centering
    \includegraphics[width=\columnwidth]{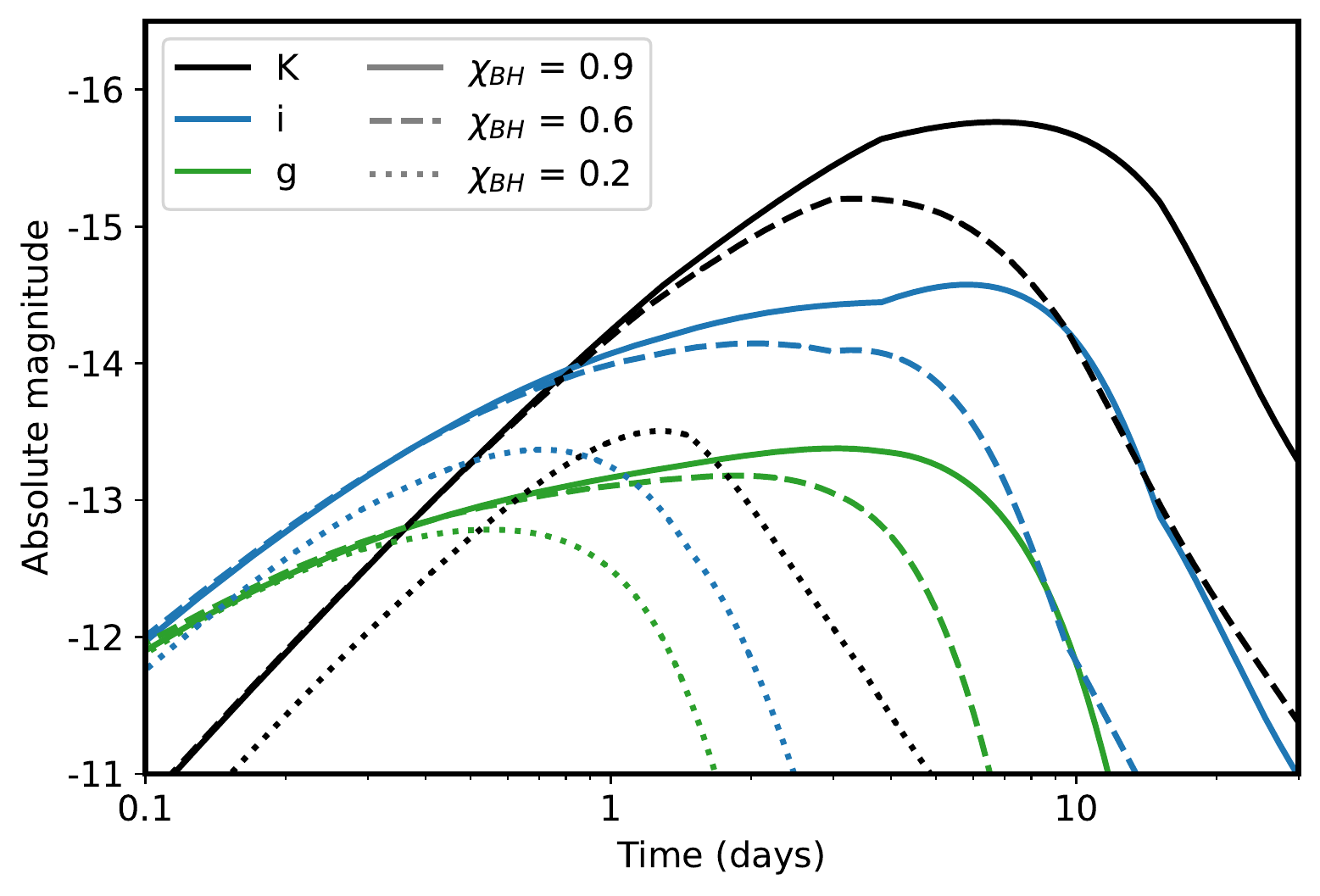}
    \includegraphics[width=\columnwidth]{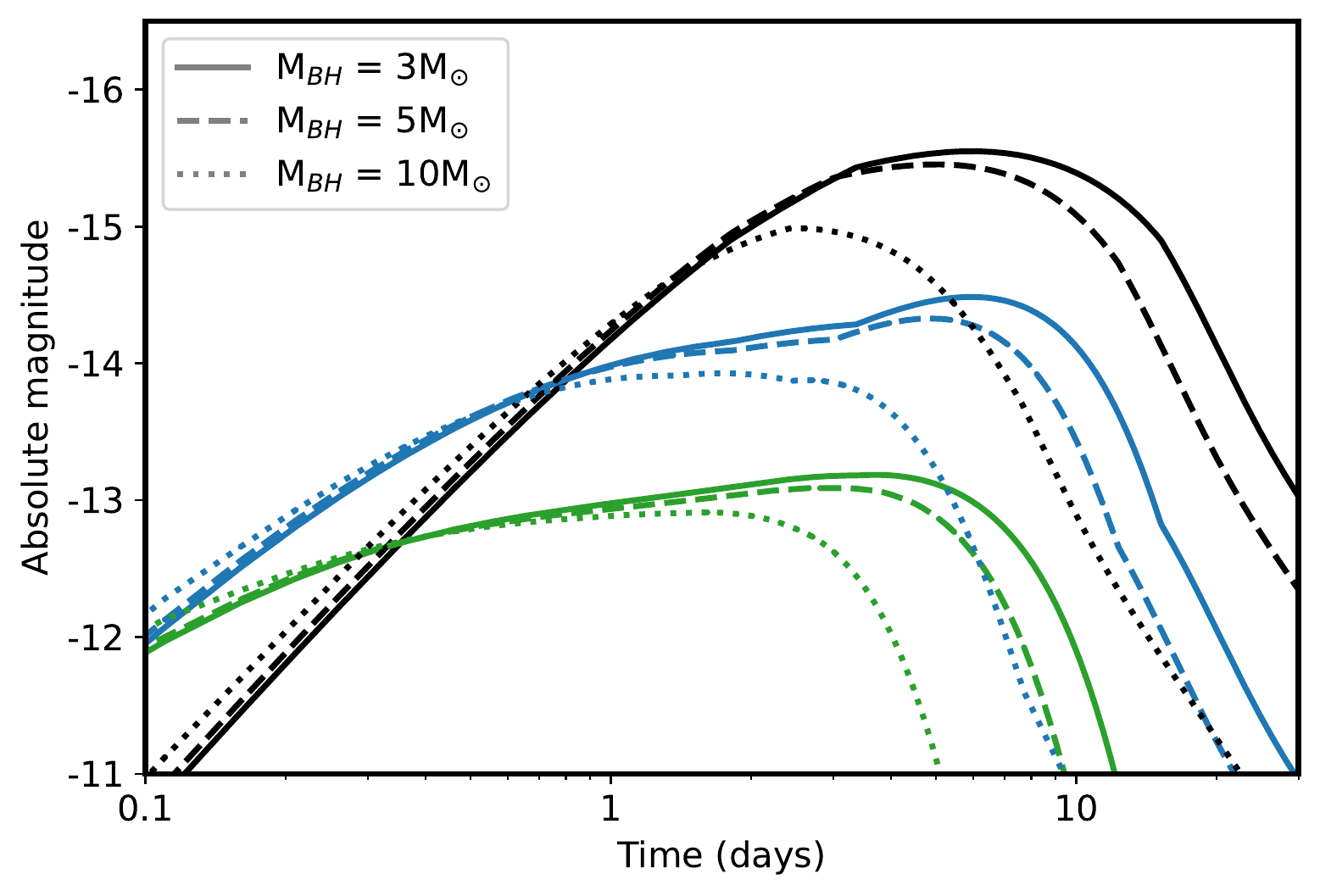}
    \includegraphics[width=\columnwidth]{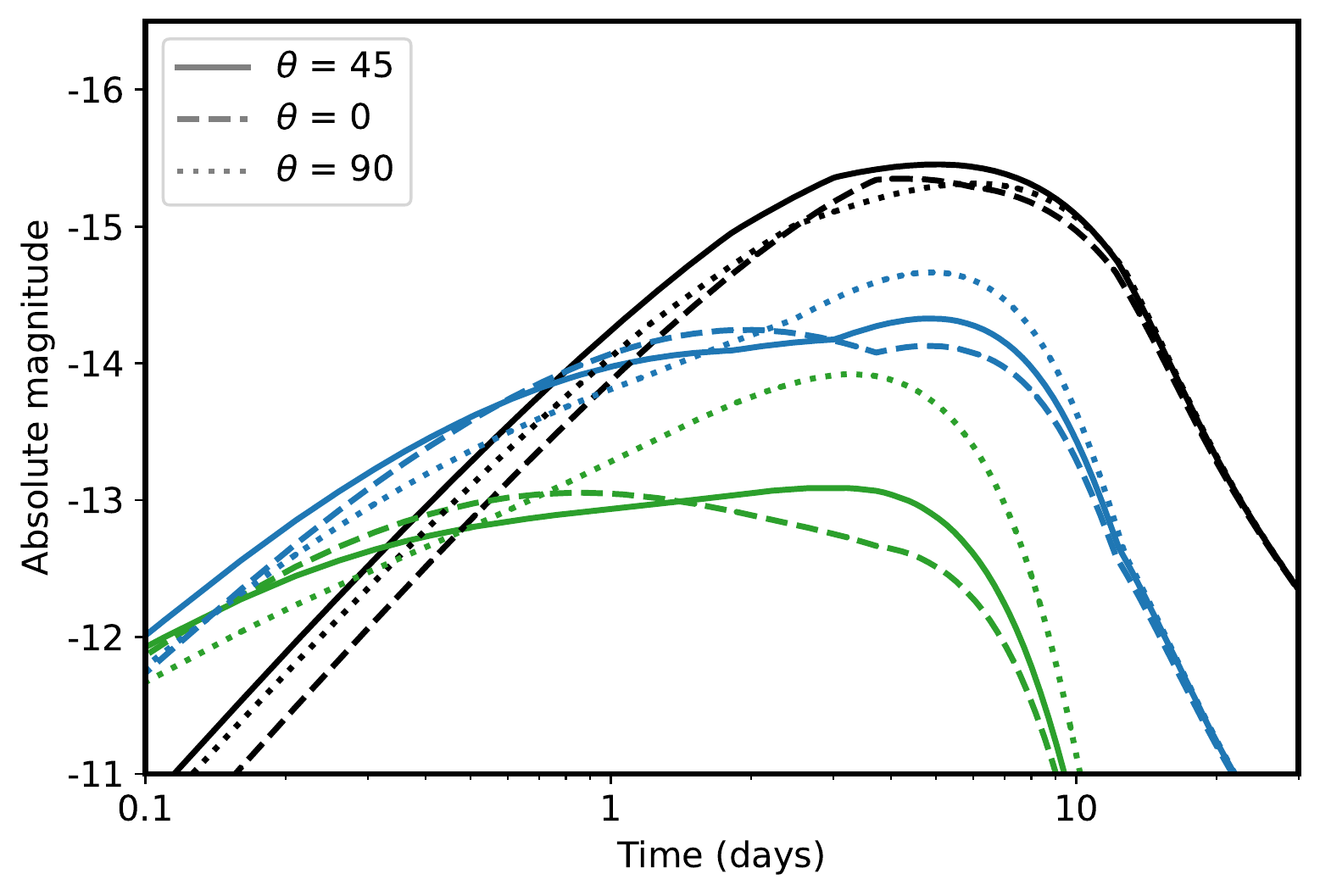}
    \includegraphics[width=\columnwidth]{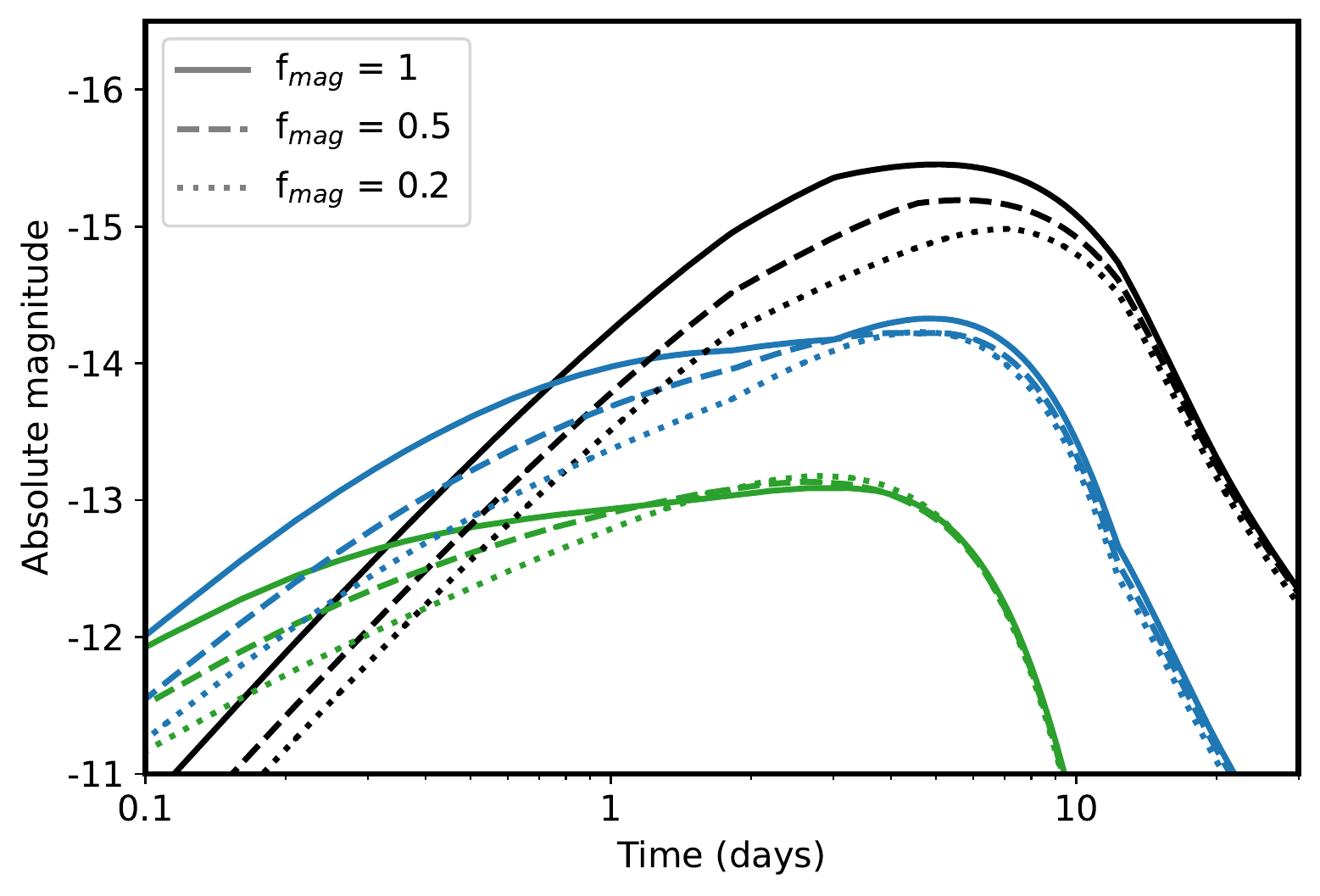}
    \caption{Example light curves in the $K$ (black), $i$ (blue), and $g$ (green) bands for our fiducial model (Table~\ref{tab:model}), with variations in a single parameter per panel. \emph{Top left:} varying $\chi_{\rm BH}$, the orbit-aligned BH spin. \emph{Top right:} varying $q$, the binary mass ratio (via $M_{\rm BH}$). \emph{Bottom left:} varying $\theta$, the observer inclination from the pole. \emph{Bottom right:} varying $f_{\rm mag}$, the magnetic field geometry.}
    \label{fig:parameters}
\end{figure*}

\subsection{Connecting to compact binary coalescences}

The shape of the kilonova light curve is underpinned by the properties of the merging binary, which can be measured from GW observations. The most accurately measured GW parameter is the `chirp' mass ($\mathcal{M}$), which is related to the binary component masses by $\mathcal{M} = (M_{\rm BH} M_{\rm NS})^{3/5} (M_{\rm BH} + M_{\rm NS})^{-1/5}$. GW measurements can also provide constraints on the viewing angle $\theta$, the mass ratio $q$ and the orbit-aligned BH dimensionless spin $\chi_{\rm BH}$.

One parameter of particular importance when estimating the mass of material that remains outside of the event horizon (Equation~\ref{eq:ejecta}) is the NS compactness, $C_{\rm NS}$. When fitting combined GW-EM multi-messenger data, $C_{\rm NS}$ can be measured rather than assumed, leading to constraints on the NS equation-of-state. From the EM side, $C_{\rm NS}$ can be constrained via the best-fit ejecta mass from the kilonova light curve. The signal detected by GW detectors is a mass-weighted combination of the tidal deformability of the two binary components, known as the effective tidal deformability \citep[$\Tilde{\Lambda}$;][]{Flanagan08,Wade14,Raithel18}. Tidal deformability is a measure of the responsiveness of a body to an external tidal field, and is zero for a BH \citep{Binnington09,Damour09}. In the NSBH case, the tidal deformability of the NS can therefore be calculated from the component masses of the binary and the effective tidal deformability:
\begin{equation}
\Lambda_{\rm NS} = \frac{13}{16} \frac{\Tilde{\Lambda}(M_{\rm BH} + M_{\rm NS})^5}{(M_{\rm NS} + 12M_{\rm BH})M_{\rm NS}^4}.
\end{equation}
We then relate this quantity to $C_{\rm NS}$ using the quasi-universal relation derived in \citet{Yagi17}:
\begin{equation}
C_{\rm NS} = 0.360 - 0.0355 \ln{(\Lambda_{\rm NS})} + 0.000705 \ln{(\Lambda_{\rm NS})}^2.
\end{equation}

Our final model consists of 9 free parameters. These are listed in Table~\ref{tab:model} with their fiducial values and assumed priors. The GW and EM branches of the model and the relationship between the measured and derived parameters is shown in Figure~\ref{fig:cartoon}.

\subsection{Parameter Sensitivity}

Figure~\ref{fig:parameters} shows how the kilonova light curves are affected by varying $\chi_{\rm BH}$, the binary mass ratio (by changing $M_{\rm BH}$), the observer angle, and the assumed dipole field configuration through $f_{\rm mag}$. As expected, higher BH spins and more symmetric binary mass ratios produce brighter kilonovae in all observing filters because they lead to a greater ejected mass outside of the remnant event horizon. We find that the $K$-band brightness is largely insensitive to viewing angle, likely due to the highly similar colour, mass, and velocity of the dynamical ejecta at the equator and the magnetically driven wind at the poles in our fiducial model. The bluer bands are more sensitive to the viewing angle, with the $g$-band light curves appearing $\sim 1.5$ magnitudes brighter at peak for an equatorial observer than a polar one at 3 -- 5 days after merger. This is because an equatorial viewing angle provides the widest range of sight lines to the thermal wind of the three viewing angles presented, and hence the largest relative contribution from the lowest opacity material to the received flux. This finding suggests that NSBH-driven kilonovae may peak quite strongly in the optical up to a week after merger for oblique viewing angles, in stark contrast to BNS events. However, significant neutrino irradiation from the inner accretion disk or preferential photon diffusion towards the poles may result in an earlier optical peak and a reversal in the observer angle dependence \citep[e.g.][]{Kawaguchi20,Zhu20}. Finally, we find that varying the magnetic field geometry has a moderate ($\sim 1$ magnitude) effect on the peak brightness in the $K$-band, due to the larger mass ejection associated with more polar field geometries (i.e. increasing $f_{\rm mag}$). A similar effect is seen in the early ($\lesssim 1$ day) optical evolution, where higher velocity magnetic winds lower the density of the ejecta more rapidly, allowing photons to escape to the observer sooner.

\section{Comparison to GRB-kilonovae}\label{sec:GRB-KNe}

\begin{figure*}
    \centering
    \includegraphics[width=\columnwidth]{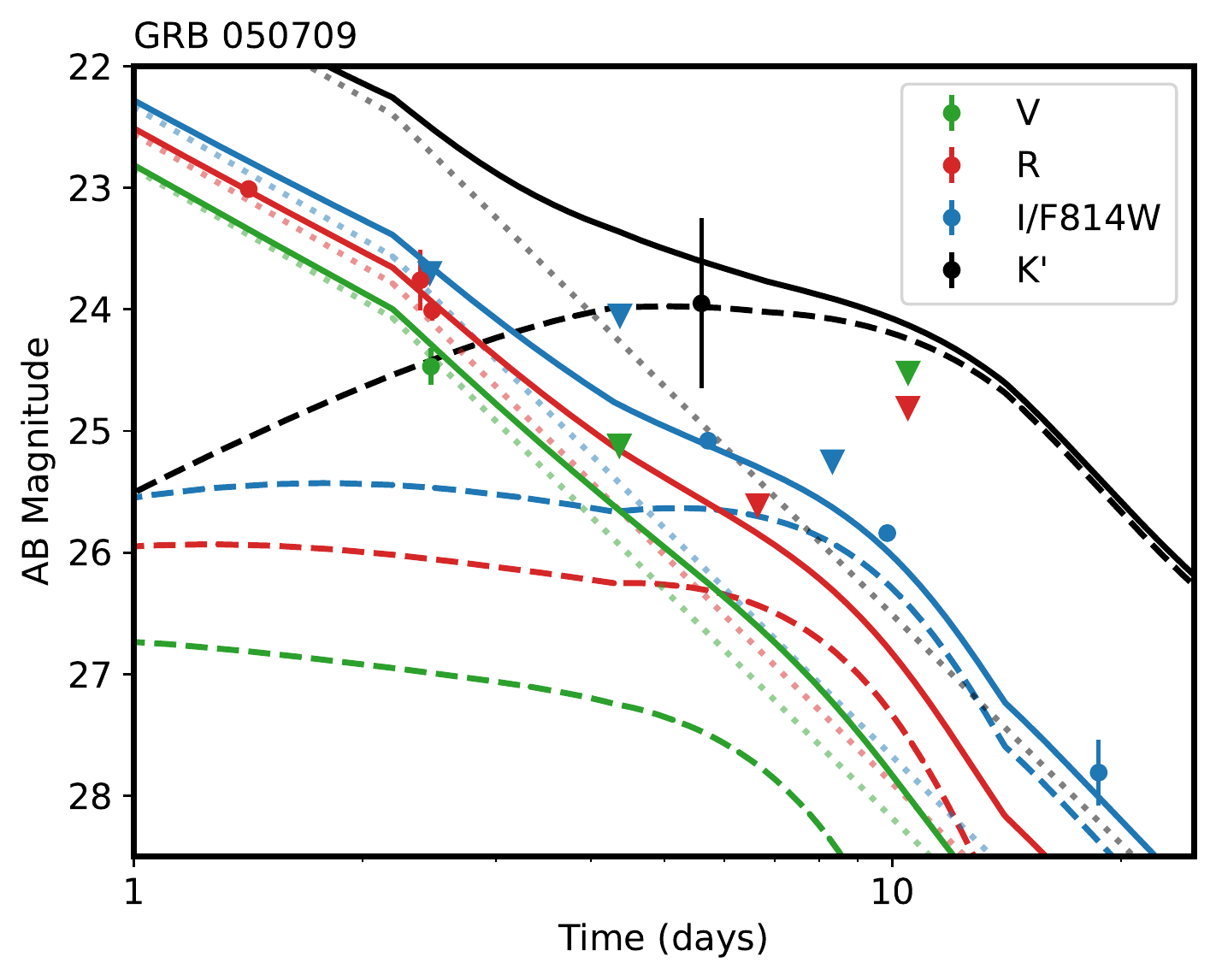}
    \includegraphics[width=\columnwidth]{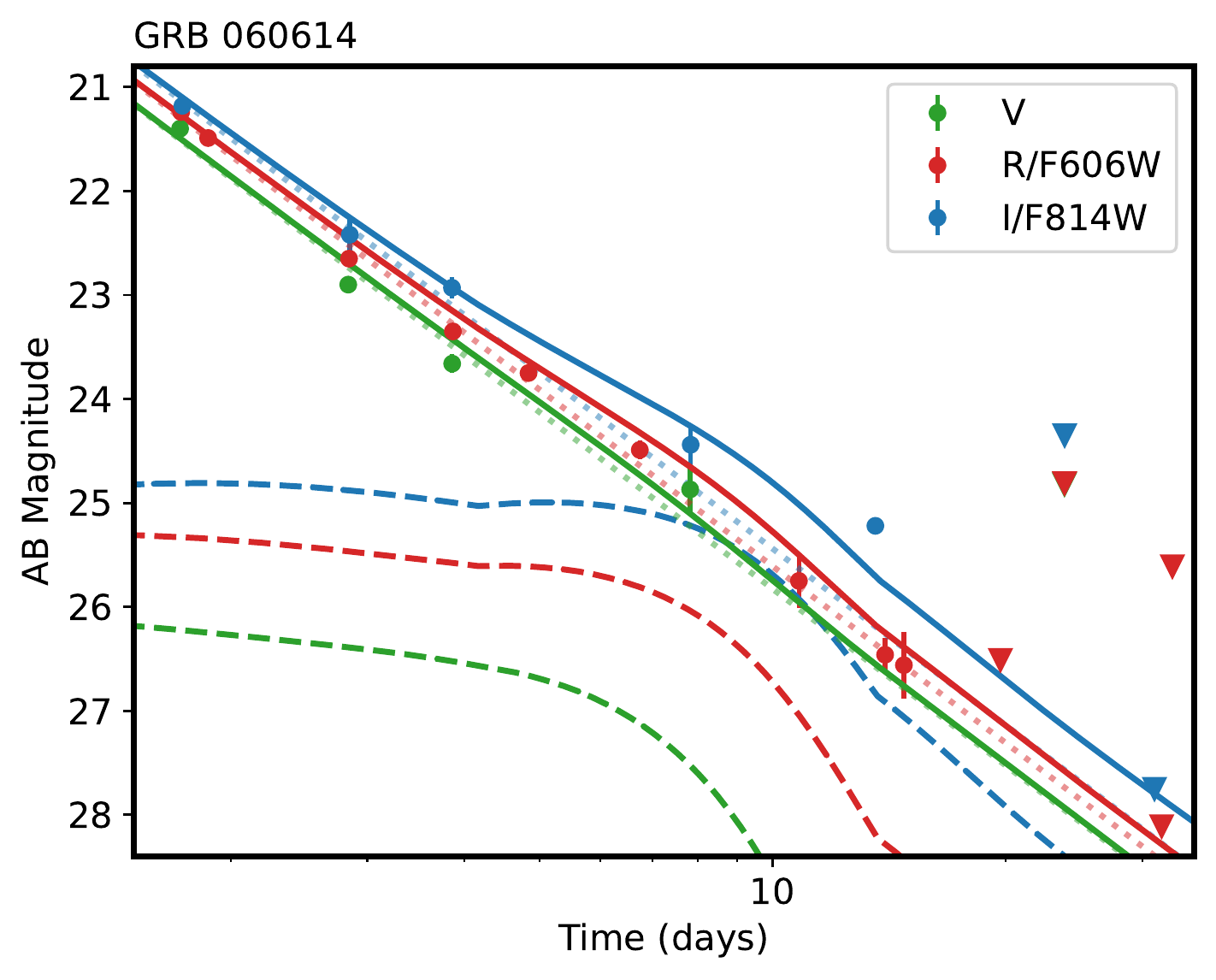}
    \includegraphics[width=\columnwidth]{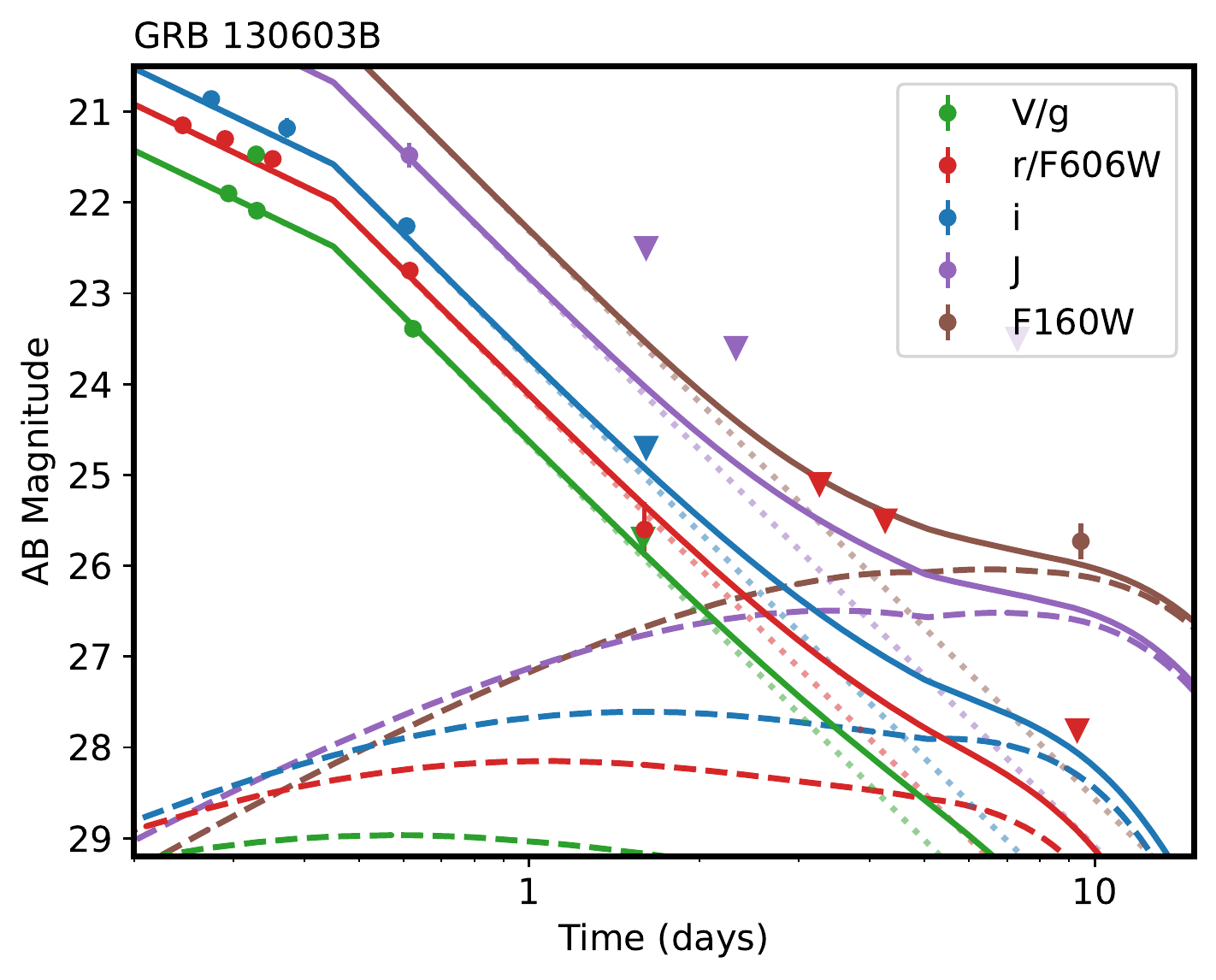}
    \includegraphics[width=\columnwidth]{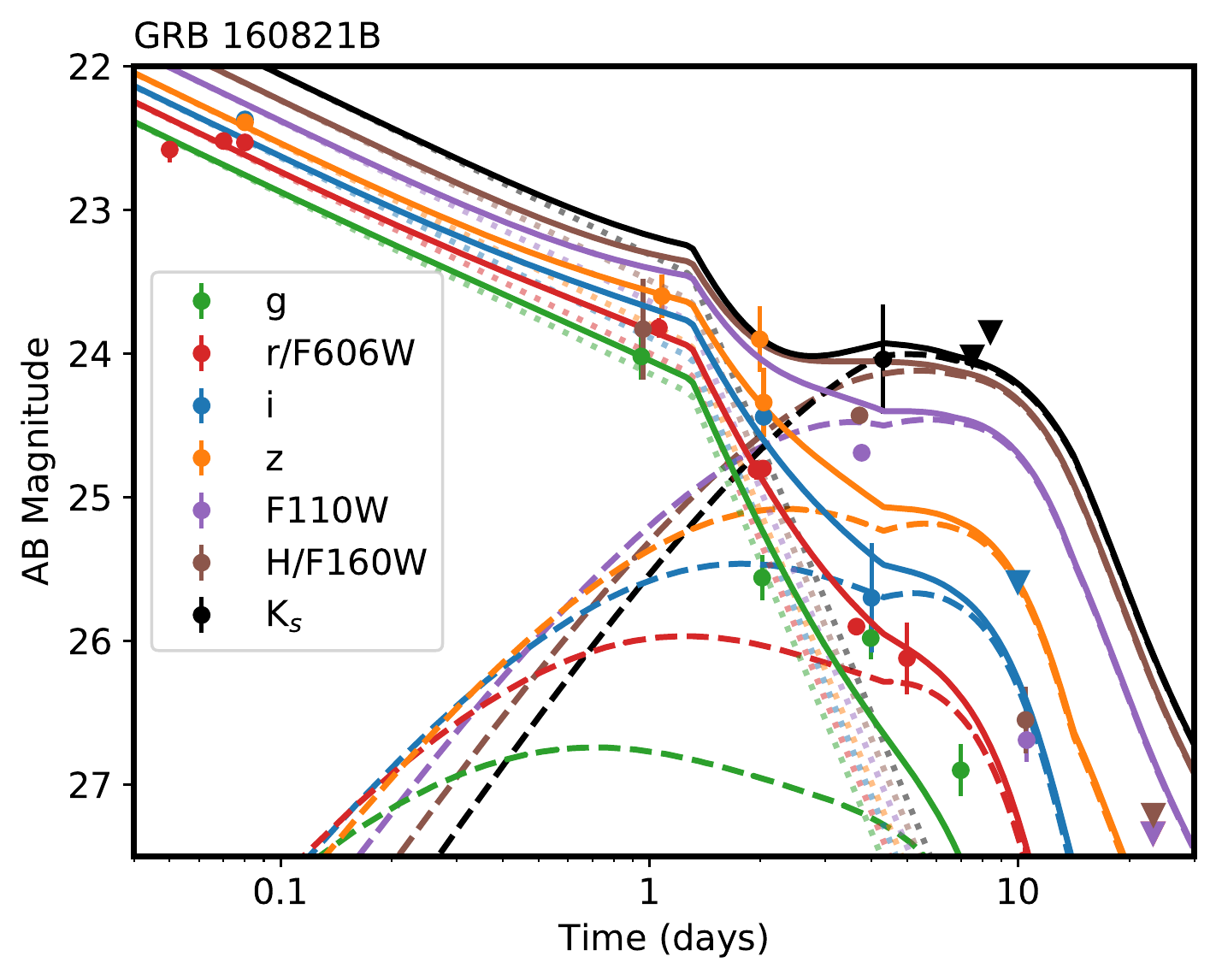}
    \includegraphics[width=\columnwidth]{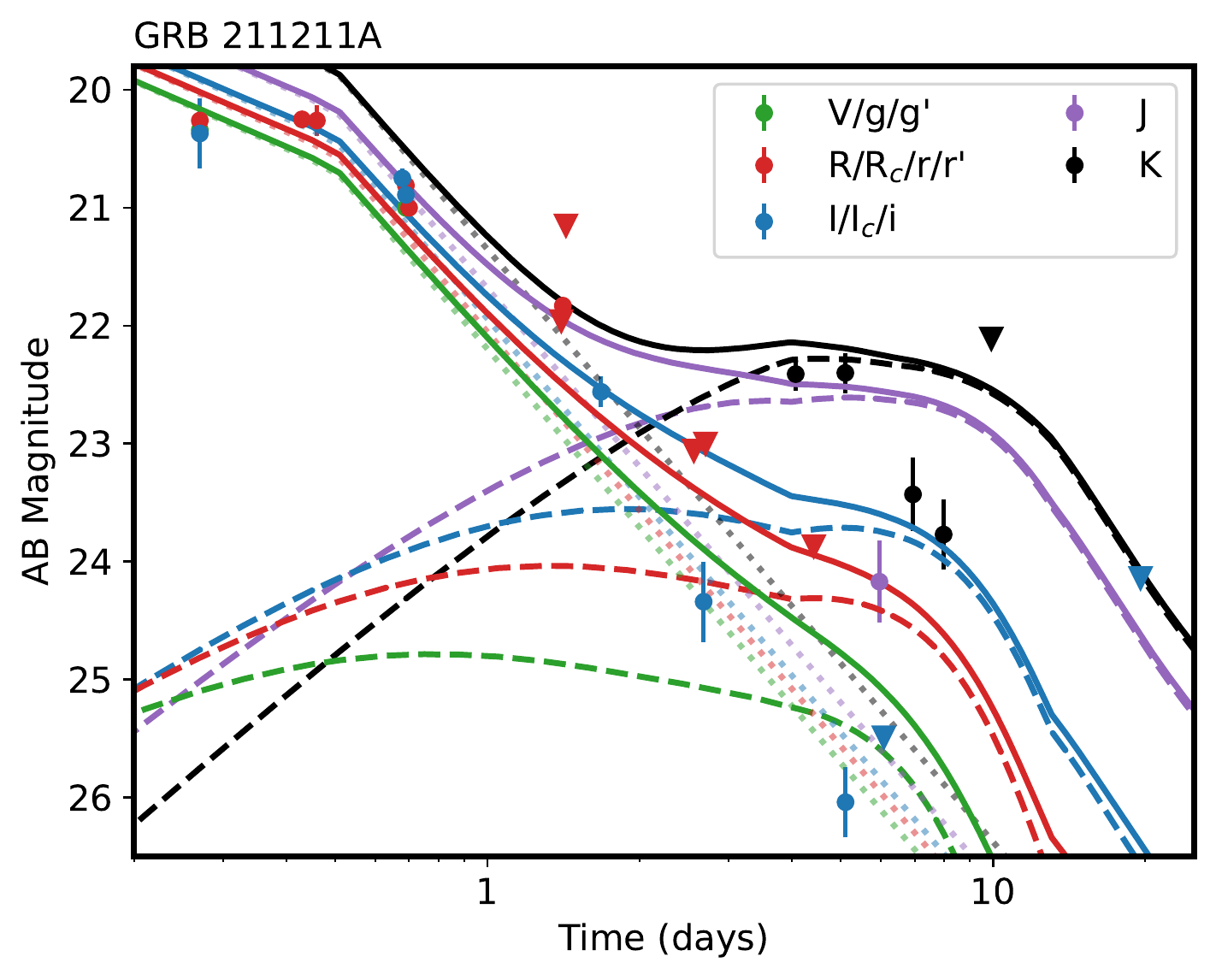}
    \caption{Model comparison to five kilonova candidates associated with cosmological GRBs. Our fiducial NSBH kilonova model (dashed lines, see Table~\ref{tab:model}) is scaled to the distance of each GRB and assumes a polar viewing angle ($\cos{\theta} = 1$). The GRB afterglow is approximated by power-law or broken power-law profiles with flux $F \propto t^{-\alpha}\nu^{-\beta}$ (dotted lines). Solid lines show the sum of the two components. Note that the kilonova models are not fit to the data in any way.}
    \label{fig:GRB_KNe}
\end{figure*}

In this section we compare a selection of kilonova candidates associated with cosmological SGRBs to our fiducial model (see Table~\ref{tab:model}). Figure~\ref{fig:GRB_KNe} shows the light curves of five afterglow + kilonova candidates. These include the first reported GRB-kilonova candidate \citep[GRB 130603B;][]{Tanvir13,Berger13}, the two best-sampled GRB-kilonovae outside of GW170817 \citep[GRB 160821B and GRB 211211A;][]{Lamb19,Troja19,Rastinejad22,Troja22,Gompertz23} and two examples of kilonova candidates alongside `extended emission' \citep[EE;][]{Norris06,Norris10,Gompertz13} SGRBs \citep[GRB 050709 and GRB 060614;][]{Yang15,Jin15,Jin16}. EE SGRBs have been suggested as candidates for NSBH-driven events \citep{Troja08,Gompertz20}, and exhibit $\sim 100$s of rapidly-evolving high energy emission \citep{Gompertz23} in addition to the $\lesssim 2$s prompt spike. In each case, our fiducial model is combined with power-law or broken-power law profiles that approximate the GRB afterglow. The parameters used are shown in Table~\ref{tab:AG_params}.

The comparisons are deliberately approximate; in many cases the available data is not sufficient to constrain the large number of parameters needed to model both the GRB afterglow and the kilonova. Nevertheless, we demonstrate that even without fine tuning, our fiducial NSBH kilonova model provides rough agreement with the candidate kilonova excesses seen in SGRBs. Our fiducial NSBH kilonova model (Table~\ref{tab:model}) produces $0.05$\,M$_{\odot}$ of dynamical ejecta (red equatorial ejecta; $\kappa_{\rm dyn} = 10$\,cm$^2$\,g$^{-1}$) with a mean velocity of 0.25c. It also produces $0.04$\,M$_{\odot}$ of magnetically-driven wind ejecta (red polar ejecta; $\kappa_{\rm mag} = 10$\,cm$^2$\,g$^{-1}$) with a mean velocity of 0.22c, and $0.04$\,M$_{\odot}$ of thermally-driven wind ejecta (``purple'' ejecta; $\kappa_{\rm mag} = 5$\,cm$^2$\,g$^{-1}$) with a mean velocity of 0.034c. We compare this to the $r$-process masses inferred using other published model fits in the literature for each GRB.

\begin{table}
    \centering
    \begin{tabular}{cccccc}
    \hline\hline
    GRB & $d_L$ & $\beta$ & $\alpha_1$ & $t_b$ & $\alpha_2$ \\
     & (Mpc) & & & (days) & \\
    \hline
    050709 & 795 & $1.0$ & $1.4$ & $2.2$ & $2.5$ \\
    060614 & 608 & $0.8$ & $2.3$ & -- & -- \\
    130603B & 1960 & $1.8$ & $1.2$ & $0.5$ & $2.5$ \\
    160821B & 806 & $0.5$ & $0.5$ & $1.3$ & $2.5$ \\
    211211A & 350 & $0.5$ & $0.8$ & $0.5$ & $2.0$ \\
    \hline\hline
    \end{tabular}
    \caption{The luminosity distance ($d_L$) of the five GRBs in Figure~\ref{fig:GRB_KNe} with the spectral ($\beta$) and temporal ($\alpha$) indices and break times ($t_b$) used to approximate their afterglows.}
    \label{tab:AG_params}
\end{table}

\subsection{GRB 050709}

GRB 050709 was detected by the High Energy Transient Explorer \citep[HETE-2;][]{Lamb00}. It featured a short, hard prompt spike with $t_{90} = 70\pm10$\,ms in the 30-400\,keV energy band, followed by a long-soft tail with $t_{90} = 130\pm7$\,s in the 2-25\,keV energy band \citep{Villasenor05}, where $t_{90}$ is the time in which the middle 90 per cent of event photons are collected. GRB 050709 is therefore an EE SGRB. It was the first SGRB for which an optical counterpart was identified \citep{Hjorth05} and was associated with a galaxy at $z = 0.16$ \citep{Fox05}. A kilonova was first claimed in GRB 050709 by \citet{Jin16}. Photometry was taken from \citet{Fox05,Covino06} and \citet{Jin16}. We find that the \citet{Jin16} $I_{\rm Vega} = 24.1 \pm 0.2$ detection at $t \sim 2.5$\,days is incompatible with the contemporaneous $g$- and $r$-band detections and preceding $r$-band detection under an afterglow interpretation, and use the $I_{\rm Vega} > 23.25$ upper limit from \citet{Covino06} for this epoch.

Our fiducial model provides a good qualitative match to the data. This is in agreement with \citet{Jin16}, who found a best fit with an ejecta mass of $0.05$\,M$_{\odot}$ and a velocity of $0.2$c from an NSBH merger, consistent with the fiducial model. However, we note that all of the data can be adequately described by a GRB afterglow model if the jet break occurs at $t \sim 10$\,days, and hence the veracity of the GRB 050709 kilonova candidate remains uncertain.

\subsection{GRB 060614}

GRB 060614 was detected by the Burst Alert Telescope \citep[BAT;][]{Barthelmy05} on board the \emph{Neil Gehrels Swift Observatory} \citep{Gehrels04}. The burst duration of $t_{90} = 102$\,s \citep[15 -- 350\,keV;][]{Gehrels06} is significantly above the canonical $t_{90} = 2$\,s divide between short and long GRBs \citep{Kouveliotou93}. However, at a redshift of $z = 0.125$ \citep{DellaValle06,Gal-Yam06}, deep optical observations exclude an associated supernova to limits hundreds of times fainter than the archetypal GRB supernova SN1998bw \citep{Fynbo06,Gal-Yam06,DellaValle06}. GRB 060614 is therefore most likely a merger-driven EE SGRB \citep[see however][]{Cobb06}, further supported by its negligible spectral lag \citep{Gehrels06} and strong spectral evolution \citep{Mangano07}. Based on its light curve, \citet{Yang15} and \citet{Jin15} claimed evidence for a kilonova counterpart. Photometry was taken from \citet{Yang15,DellaValle06} and \citet{Gal-Yam06}.

The emission of GRB 060614 is likely dominated by the bright afterglow at almost all epochs; a deviation from a power-law is only detected in two points \citep{Yang15}. Our fiducial model provides a reasonable approximation of the $i$-band excess at $\approx 8$ days, but under-predicts the flux in the $\approx 13$ day epoch. Fine tuning to produce a slightly fainter and longer-lived kilonova signature may resolve the discrepancy, which can be achieved with e.g. a lower velocity wind. \citet{Yang15} suggest an NSBH merger with kilonova ejecta mass of $\approx 0.1$\,M$_{\odot}$ and velocity $\approx 0.2c$, with an effective temperature of $\approx 2000$\,K. This is broadly consistent with the fiducial model, which produces $\approx 0.07$\,M$_{\odot}$ of ejecta combined between the magnetic and thermal winds, at a temperature of $2500$\,K and $v_{\rm mag} = 0.22c$.

\subsection{GRB 130603B}

GRB 130603B was detected by \emph{Swift}-BAT with a duration of $t_{90} = 0.18 \pm 0.02$\,s \citep[15 -- 350\,keV;][]{Lien16} and is therefore an unequivocal member of the SGRB class. With a redshift of $z = 0.356$ \citep{Cucchiara13,Thone13}, it is also the most distant GRB in our comparison sample. GRB 130603B was the first ever identified kilonova candidate \citep{Tanvir13,Berger13} thanks to a significant excess in HST F160W over the expected afterglow, constrained by a simultaneous HST F606W non-detection. Photometry was taken from \citet{Tanvir13}.

The fiducial model provides a good match to the data, although the kilonova is only detected in a single epoch and hence the observations are not particularly constraining. \citet{Kawaguchi16} showed that GRB 130603B can be described with a NSBH-driven kilonova model for reasonably high spins ($\chi_{\rm BH} > 0.3$) and larger NS radii. \citet{Berger13} find that the light curve can be described by a kilonova driven by either a BNS or BHNS with an ejecta mass of $0.03$ -- $0.08$\,M$_{\odot}$ and a velocity in the range of $0.1$ -- $0.3$c, consistent with our fiducial model. \citet{Tanvir13} find a similar mass range: $10^{-3}$\,M$_{\odot} < M_{\rm ej} < 10^{-2}$\,M$_{\odot}$.

\subsection{GRB 160821B}

GRB 160821B was detected by \emph{Swift}-BAT with $t_{90} = 0.48 \pm 0.07$s \citep[15 -- 350\,keV;][]{Lien16}. The kilonova was reported independently by \citet{Lamb19} and \citet{Troja19}, with the redshift found to be $z = 0.16$. Multi-wavelength observations, particularly those at X-ray and radio frequencies, suggested that GRB 160821B afterglow may have experienced late energy injection from a second blast wave arriving at the afterglow emission site at late times \citep{Lamb19}. Such a phenomenon is not captured in our simple power-law representation of the afterglow. We use the photometry from \citet{Kasliwal17} and \citet{Lamb19}.

Despite higher sampling than most of the other GRBs presented in this work, the fiducial model does remarkably well in matching the evolution of GRB 160821B with no fine tuning of the kilonova. We note that the $J$- and $H$-bands are over-predicted, particularly at late times, implying that the mass of the reddest ejecta needs to be reduced or its emission evolve faster. This can be achieved with a lower binary mass ratio or BH spin. We also under-predict the emission in the $g$-band, which may indicate a lower grey opacity or higher blue ejecta fraction (from the thermal wind) is needed. \citet{Lamb19} find a good fit to the data with a refreshed afterglow and a two-component kilonova model with a wind ejecta mass of $0.01$\,M$_{\odot}$ travelling at $v < 0.15$c and a dynamical ejecta mass of $0.001$\,M$_{\odot}$ with $v > 0.1$c. \citet{Troja19} find a low, lanthanide-rich ($\kappa = 10$\,cm$^2$\,g$^{-1}$) ejecta mass of $\lesssim 0.006$\,M$_{\odot}$ and $v \gtrsim 0.05$c. The low ejecta masses inferred by both studies are as much as an order of magnitude less than is produced in our fiducial model, and may explain why it over-predicts the late near-infrared evolution.

\subsection{GRB 211211A}

GRB 211211A was detected by the \emph{Fermi} Gamma-ray Burst Monitor \citep[GBM;][]{Meegan09} and \emph{Swift}-BAT, with the latter measuring $t_{90} = 51.4 \pm 0.8$\,s \citep[15 -- 350\,keV;][]{Stamatikos21}. The burst is therefore an EE SGRB \citep[for a full analysis of the high energy emission see][]{Gompertz23}. At a redshift of $z = 0.076$ \citep{Rastinejad22}, GRB 211211A is the second-closest compact binary merger to Earth ever discovered, with only the GW-localised GW170817 more proximal. The kilonova was identified through a strong infrared excess by \citet{Rastinejad22} and was independently modelled by \citet{Mei22,Troja22,Xiao22,Yang22} and \citet{Zhu22}. We use the photometry from \citet{Rastinejad22}.

Similar to GRB 160821B, our fiducial model struggles to evolve fast enough to reproduce the late near-infrared observations. It over-predicts the flux at essentially all wavelengths beyond $\sim 2$ days, particularly in the $i$-band \citep[though we note that these suffer from significant systematic errors in their magnitude measurments;][]{Rastinejad22}. \citet{Rastinejad22} find a best fit kilonova model with a total ejecta mass of $0.047^{+0.026}_{-0.011}$\,M$_{\odot}$, half of which is partitioned in a lanthanide-rich `red' component with $v \approx 0.3$c. The other half is divided equally between an intermediate-opacity `purple' component with $v \approx 0.1$c and a lanthanide-free `blue' component with $v \approx 0.3$c. A BNS merger was preferred over an NSBH. The total ejecta mass is lower than is produced by the fiducial model. The relative abundance of lanthanide-rich, high-velocity ejecta could be achieved with a strong magnetic wind in our model, implying a high magnetic field with poloidal geometry. A strong dipole field is inferred for GRB 211211A by \citet{Gao22}.

\citet{Mei22} fit the observations with an isotropic, one-component kilonova model. They find an ejecta mass of $0.020^{+0.009}_{-0.006}$\,M$_{\odot}$ with an average velocity of $0.10^{+0.07}_{-0.04}c$ and a grey opacity of $0.6^{+0.8}_{-0.3}$\,cm$^2$\,g$^{-1}$. \citet{Troja22} find that the observations can be matched with $0.01$ -- $0.1$\,M$_{\odot}$ of wind ejecta and $0.01$ -- $0.03$\,M$_{\odot}$ of dynamical ejecta from a BNS merger. \citet{Zhu22} employ an NSBH binary-driven model, and find the observations are best described by the merger of a $8.21^{+0.77}_{-0.75}$\,M$_{\odot}$ BH with dimensionless spin $0.62^{+0.06}_{-0.07}$ with a $1.23^{+0.06}_{-0.07}$\,M$_{\odot}$ NS, producing $0.005$ -- $0.03$\,M$_{\odot}$ of lanthanide-poor wind ejecta and $0.015$ -- $0.025$\,M$_{\odot}$ of lanthanide-rich dynamical ejecta. Finally, \citet{Yang22} find a lanthanide-poor kilonova ($\kappa = 0.8^{+0.1}_{-0.2}$\,cm$^2$\,g$^{-1}$) with a total ejecta mass of $0.027^{+0.011}_{-0.001}$\,M$_{\odot}$ at an average velocity of $0.25^{+0.06}_{-0.02}$c. 

\section{Fitting to GW 170817}\label{sec:170817}

There is only one GW-EM multi-messenger dataset available for fitting: that of GW 170817 \citep{Abbott17_multimessenger}. Observational and modelling evidence strongly supports this event being a BNS merger, but some parameter space is available for NSBH models with BH masses below typical expectations. \citet{Coughlin19b} showed that an NSBH merger could potentially reproduce the GW and EM constraints, but is disfavoured relative to a BNS merger.

In this section we investigate the ability of our model to reproduce the GW signal and kilonova associated with GW170817. We approach this in two different ways. In the first approach, we use an `astrophysical' prior that does not include the posteriors derived from GW170817, and instead allows the model to explore the full parameter space for NSBH mergers that are expected to be EM bright while penalising realisations that lie outside of theoretical expectations. Specifically, we penalise solutions with NSs more massive than the maximum stable NS mass \citep[the Tolman-Oppenheimer-Volkoff mass, $M_{\rm TOV}$, e.g.][]{Shapiro86}, which we set as $M_{\rm TOV} = 2.17$ \citep{Margalit17, Nicholl21}, as well as BHs with masses below this threshold. We also penalise solutions with tidal deformabilities outside of the expected range \citep[e.g.][]{Hinderer08,Hinderer10,Postnikov10}. The second approach takes the posterior solutions from \citet{Abbott17_GW} as the model priors, but relaxes the penalties for unconventional solutions. The first formalism therefore allows the model to search for a more `canonical' NSBH binary system that best reproduces the light curve, and the second challenges it to find a solution that satisfies the GW signal even where it defies expectations.

\begin{figure*}
\centering
    \includegraphics[width=\columnwidth]{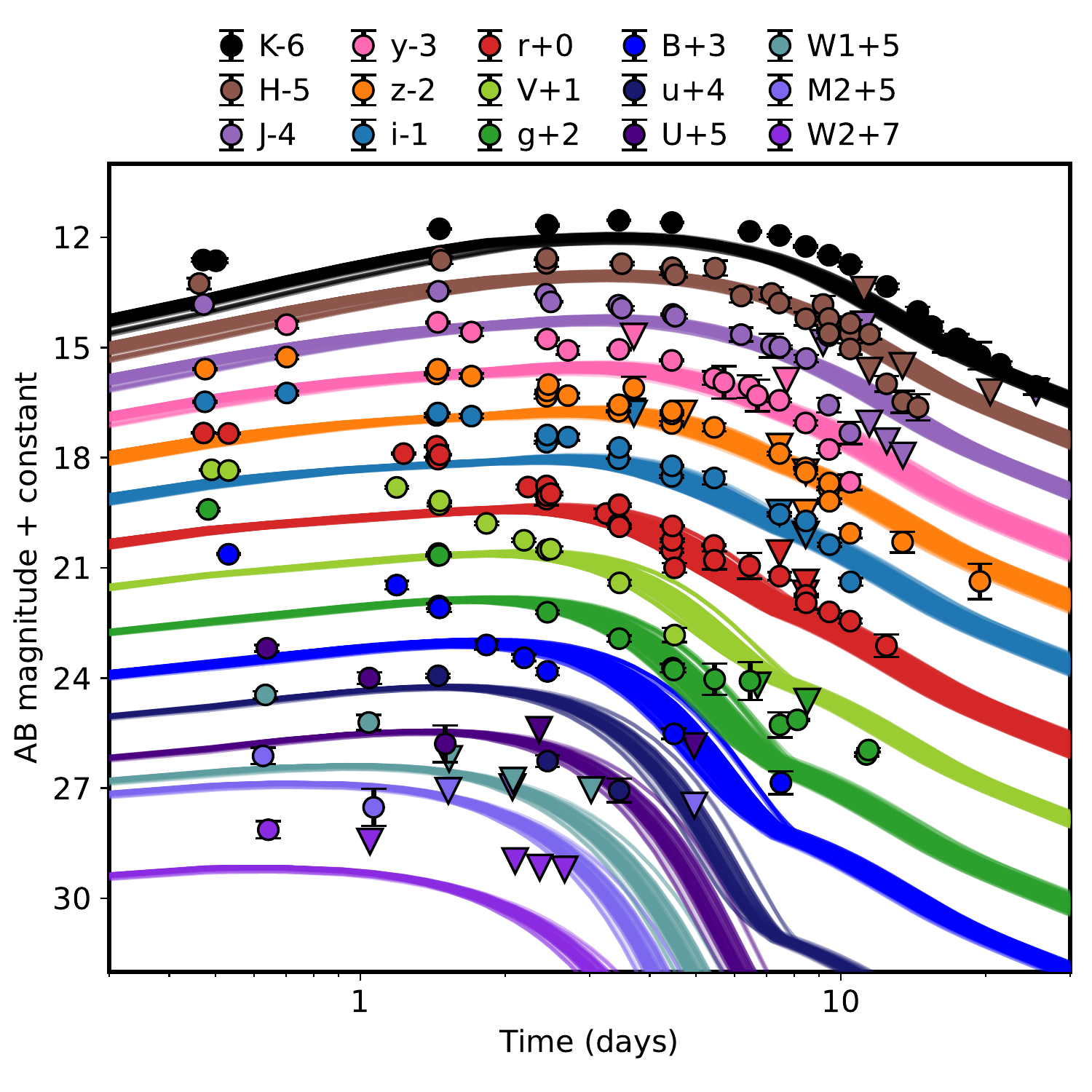}
    \includegraphics[width=\columnwidth]{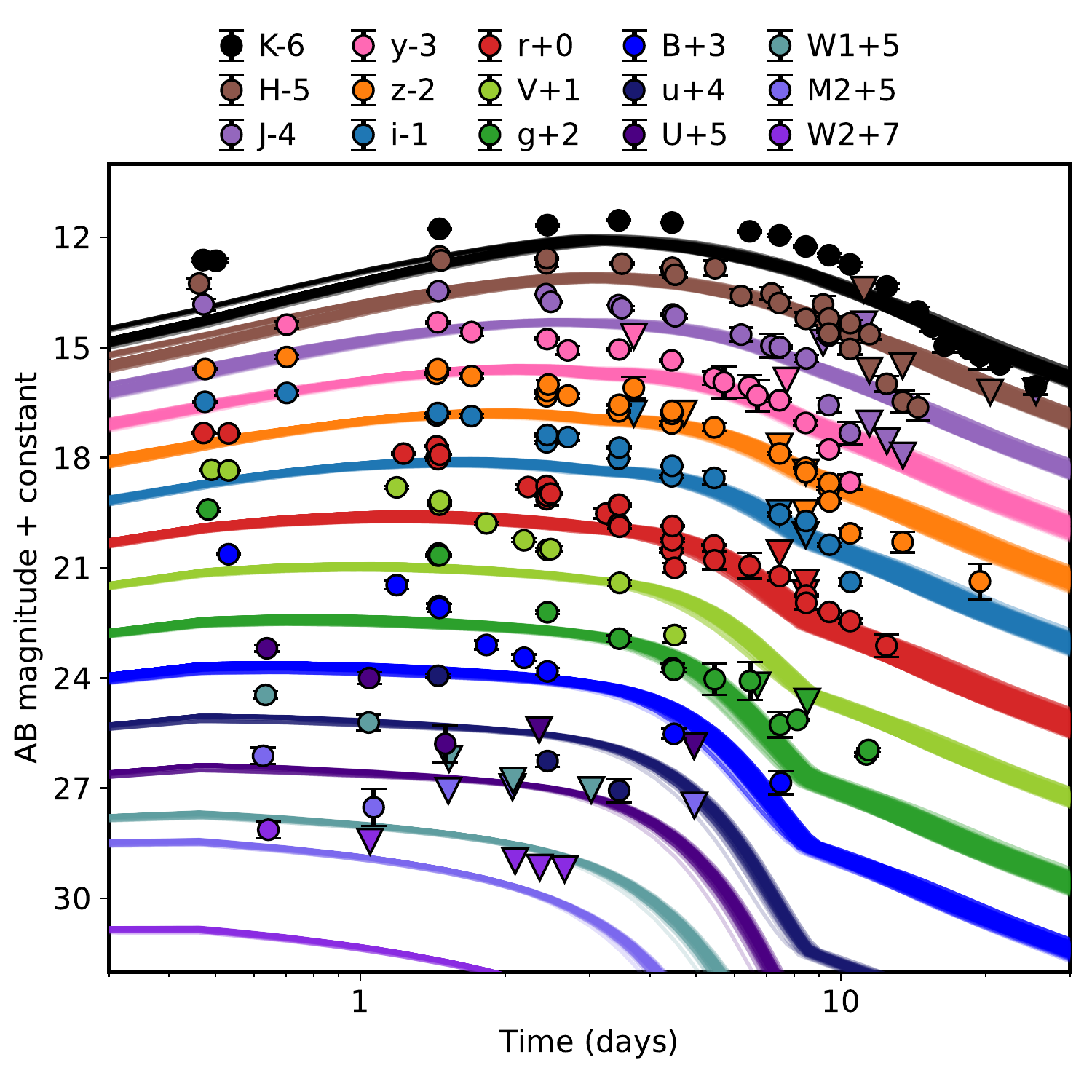}
    \caption{Light curves from the posterior distributions of the best fits to the GW170817 data \citep{Villar17} using the NSBH kilonova model and the general astrophysical priors (left) or event-based GW170817 priors (right). The model provides a reasonable match to the data at times later than two days after trigger, but struggles to produce the early emission in both cases, particularly in optical and UV bands.}
    \label{fig:GW170817}
\end{figure*}

Fitting is performed with {\sc emcee} \citep{Foreman-Mackey13}. Our best-fitting solutions for the two prior sets are shown in Figure~\ref{fig:GW170817}. While the model provides a reasonably good match to the late emission and redder bands, it significantly under-produces the early optical emission. This result is expected; even BNS models, which are capable of providing more `blue' emission than the NSBH case, require additional means for producing optical light when modelling GW170817 and other well-sampled kilonovae \citep{Nicholl21,Rastinejad22}. Whether additional emissive mechanisms such as the shock heating of ejecta by a GRB jet \citep[e.g.][]{Kasliwal17b,Arcavi18,Piro18} can be included in NSBH models depends on whether sufficient polar material is present prior to the launching of the jet (if one is launched at all by NSBH mergers), and will require GW-EM observations to confirm. It is notable that beyond $\sim 2$ days, it becomes very difficult to distinguish the light curves of kilonovae driven by BNSs and NSBHs, and hence early observations are essential where the GW signal is either absent or ambiguous.

The posteriors for the astrophysical prior show a loose preference for a chirp mass of $\mathcal{M} = 3.01^{+0.70}_{-0.56}$, a mass ratio of $q = 0.11^{+0.03}_{-0.00}$, and an effective tidal deformability of $\tilde{\Lambda} = 0.91^{+0.88}_{-0.23}$. This translates into $M_{\rm BH} \approx 11.6$\,M$_{\odot}$, $M_{\rm NS} \approx 1.3$\,M$_{\odot}$, and $R_{\rm NS} \approx 12.1$\,km. The BH spin is preferentially high, at $\chi_{\rm BH} = 0.82^{+0.04}_{-0.05}$. The viewing angle is equatorial, $\cos{\theta} = 0.07^{+0.09}_{-0.06}$, and the magnetic field geometry is strongly dipolar at $f_{\rm mag} = 0.94^{+0.05}_{-0.21}$. Broadly, these parameters maximise the ejected mass while retaining sight lines to the bluer material.

The event-based priors limit the posterior solutions to within $28^{\circ}$ of the poles \citep{Abbott17_GW}, leading to a polar solution with $\cos{\theta} = 1.00^{+0.00}_{-0.01}$, in contrast to the results from the less restrictive astrophysical prior set. The preferred chirp mass is $\mathcal{M} = 1.19^{+0.00}_{-0.00}$, strongly constrained by the tight Gaussian priors from the GW detection. The binary mass ratio is found to be $q = 0.41^{+0.01}_{-0.01}$, with an effective tidal deformability of $\tilde{\Lambda} = 122.1^{+16.0}_{-9.1}$. These properties define a binary with $M_{\rm BH} \approx 2.2$\,M$_{\odot}$, $M_{\rm NS} \approx 0.9$\,M$_{\odot}$, and $R_{\rm NS} \approx 9.7$\,km.

The low component masses for the event-based priors are dictated by the tight constraints on the chirp mass from GW170817. This has a knock-on effect of requiring a small NS radius to avoid over-producing the emission; the NS tidal deformability is already $\sim 1500$ with $R_{\rm NS} \approx 9.7$\,km. However, this combination of low NS mass and radius would point to very stiff equations of state. The BH is found to have a relatively low orbit-aligned spin magnitude, with $\chi_{\rm BH} = 0.15^{+0.02}_{-0.04}$, again mandated by the GW priors. As with the astrophysical prior set, the magnetic field geometry is preferentially dipolar (the configuration that produces the most ejecta mass, and hence luminosity), with $f_{\rm mag} = 0.99^{+0.01}_{-0.04}$. While the binary solutions are notably different between the two prior sets, the resultant kilonovae are strikingly similar (Figure~\ref{fig:GW170817}). These results suggest that the biggest discriminant of merger type comes from the bluer bands, where bright emission from high $Y_e$ dynamical ejecta driven from the poles is produced at early times ($\lesssim$ 2 days) in the BNS model but not in our NSBH model. However, blue emission may be produced in the polar outflows of an NSBH merger if the neutrino flux from the remnant disk can raise the electron fraction of the ejecta sufficiently \citep[e.g.][]{Zhu20}, or if photons preferentially diffuse to polar regions due to high equatorial opacities \citep[e.g.][]{Kawaguchi20}. Concurrent optical and nIR monitoring will be essential to distinguish between these possibilities.

\section{Detectability with Rubin}\label{sec:LSST}

\begin{figure*}
    \centering
    \includegraphics[width=\columnwidth]{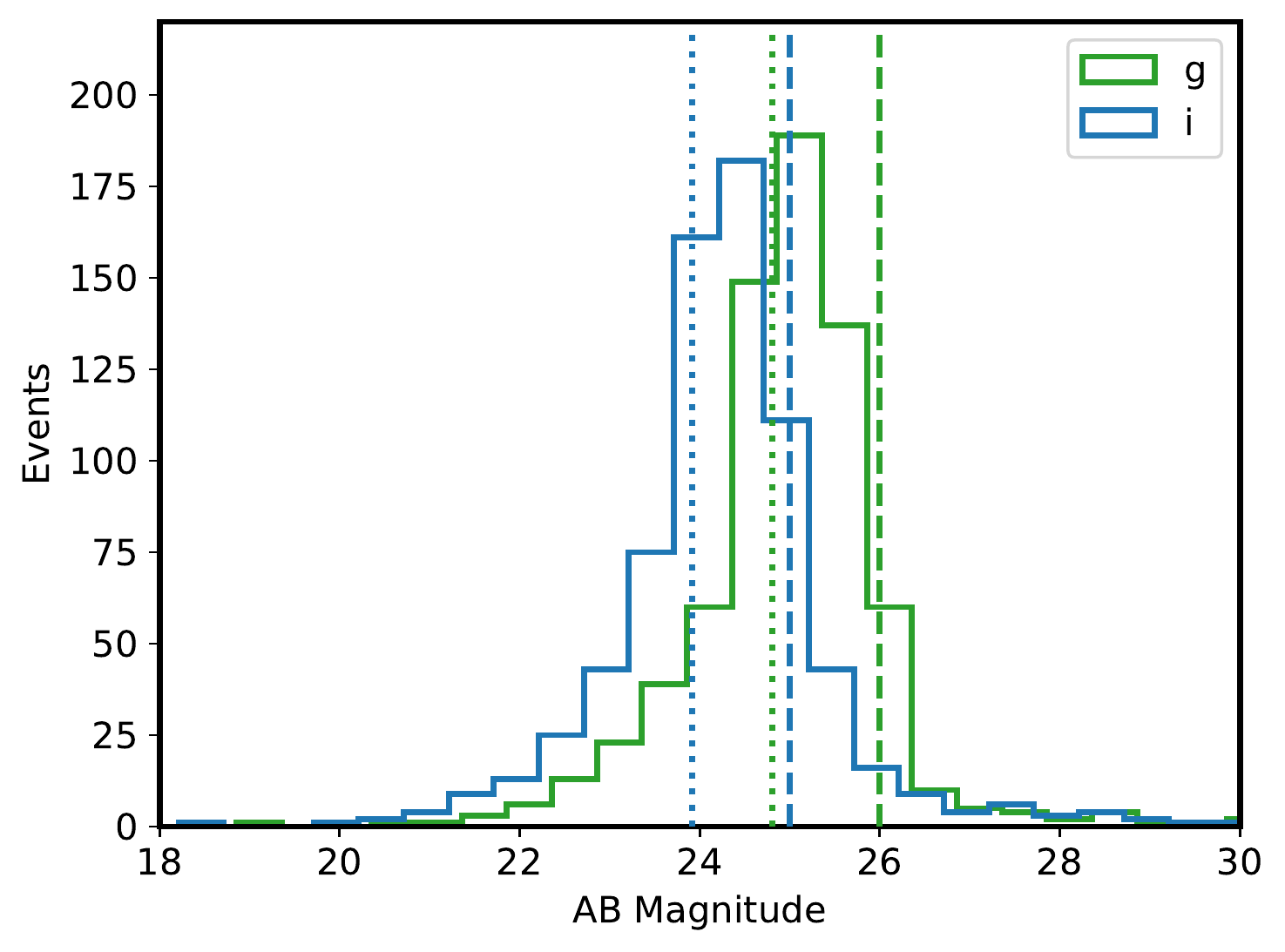}
    \includegraphics[width=\columnwidth]{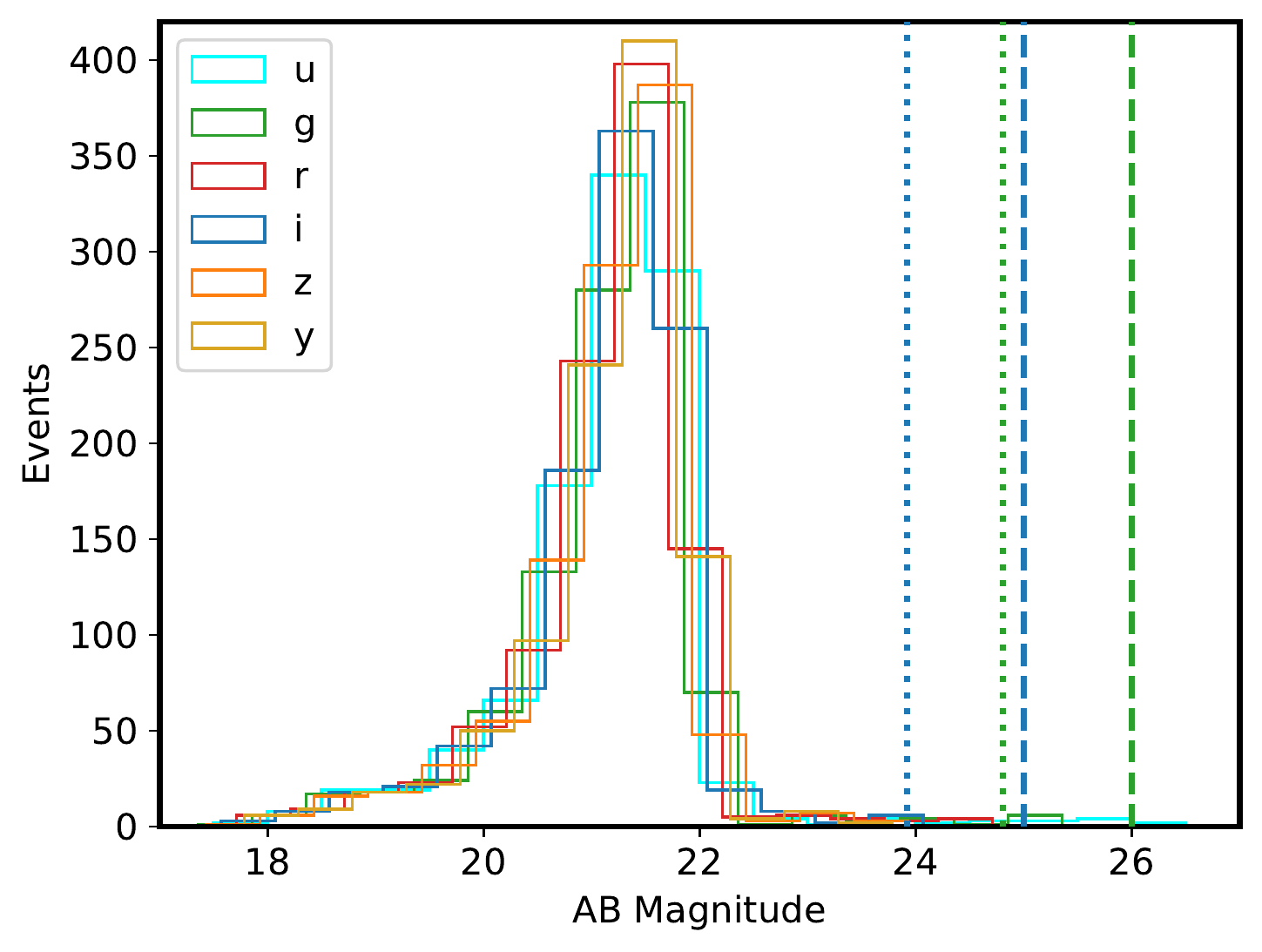}
    \caption{Histograms of the peak kilonova magnitude of 2000 realisations from our NSBH model (left) and the BNS model of \citet{Nicholl21} (right) in different photometric filters. The dotted lines show the single visit LSST WFD survey limits \citep{Ivezic19} for the g and i bands, and the GW follow-up limits \citep{Andreoni22} are shown by the dashed lines.}
    \label{fig:detectability}
\end{figure*}

With its wide field of view and large aperture, the Vera C. Rubin Observatory (Rubin) will be well suited to discovering EM counterparts to GW triggers \citep[e.g.][]{Andreoni22} and serendipitous transients during its Legacy Survey of Space and Time \citep[LSST;][]{Ivezic19}. We investigate the detectability of the population of NSBH kilonovae predicted by our model by drawing 2000 light curve realisations from our model in generative mode. This mode enables the user to define their priors in terms of component mass and NS radius rather than chirp mass and deformability, and hence is suitable for simulating populations or fitting when no GW data are available. In particular, it helps to avoid realisations with unrealistic NS masses, which are hard to mitigate against when defining populations with chirp mass and mass ratio. BH and NS mass prior distributions were constructed following model C in \citet{Broekgaarden21}, and the NS radius was set to 11\,km, following \citet{Nicholl21}. We generate our population out to 600\,Mpc from Earth, which covers the full NSBH detection range predicted for advanced LIGO in O5 \citep{Abbott20}. Priors for the other parameters were taken from the astrophysical set (Table~\ref{tab:model}), but negative $\chi_{\rm BH}$ was excluded.

Of the 2000 realisations, 727 produce detectable emission, 1273 are EM dark, and 8 were discarded for numerical artefacts in their light curves. Histograms of the peak magnitudes of our realisations in different bands are shown in Figure~\ref{fig:detectability}. Assuming a detection threshold of $g = 24.81$ and $i = 23.92$ for LSST's Wide Fast Deep (WFD) survey \citep{Ivezic19}, we obtain 279 g-band detections and 237 i-band detections. Assuming a GW follow-up strategy that reaches a depth of $g = 26$ and $i = 25$ \citep{Andreoni22}, this becomes 655 g-band detections and 599 i-band detections. However, we define `detections' as realisations with peak magnitudes above the detection threshold. In reality, it is unlikely these transients would be recovered from faint detections in single epochs. We also do not account for line-of-sight extinction and the cadence of follow-up observations. These estimates of the fraction of realisations that are detectable should therefore be considered upper limits.

Our model therefore predicts that less than one third of NSBH GW triggers (assuming a maximum distance of 600\,Mpc from Earth) will yield EM detections with LSST even if all are well sampled by follow-up observations. Sampling the whole localisation region on each of the first four nights following a GW trigger, as per the \emph{preferred} strategy of \citet{Andreoni22}, would achieve sufficient coverage. By contrast, the reduced depth of the WFD survey compared to GW follow-up observations makes it significantly less likely that our model kilonovae would be serendipitously discovered. Only $\sim 10$ per cent of realisations would be detectable if observed by chance at peak magnitude.

By way of comparison, we also investigate the detectability of a population of 2000 BNS mergers, using the model of \citet{Nicholl21}. NS mass and mass ratio priors are chosen to be Gaussian and are taken from the sample of Galactic BNS systems presented in \citet{Farrow19}. We limit our population distance to 300\,Mpc, in accordance with the advanced LIGO range for BNS systems. We find that essentially every realisation peaks at least one magnitude brighter than even the shallower detection threshold (Figure~\ref{fig:detectability}), and hence we conclude that Rubin will be capable of finding all EM counterparts to BNS GW triggers if it responds to them within a few days. However, like in the NSBH case, we neglect line-of-sight extinction and the observing cadence, defining detections only by the brightness of the realisation relative to the detectability threshold. We note that the brightness of EM counterparts (and hence their detectability) may be enhanced by gravitational lensing. Evidence of this may manifest in the GW signals, in particular in candidate `mass gap' mergers where one (or both) binary constituents are placed in the range 3 -- 5\,$M_{\odot}$ in low latency \citep{Smith23}.

\begin{figure*}
    \centering
    \includegraphics[width=\columnwidth]{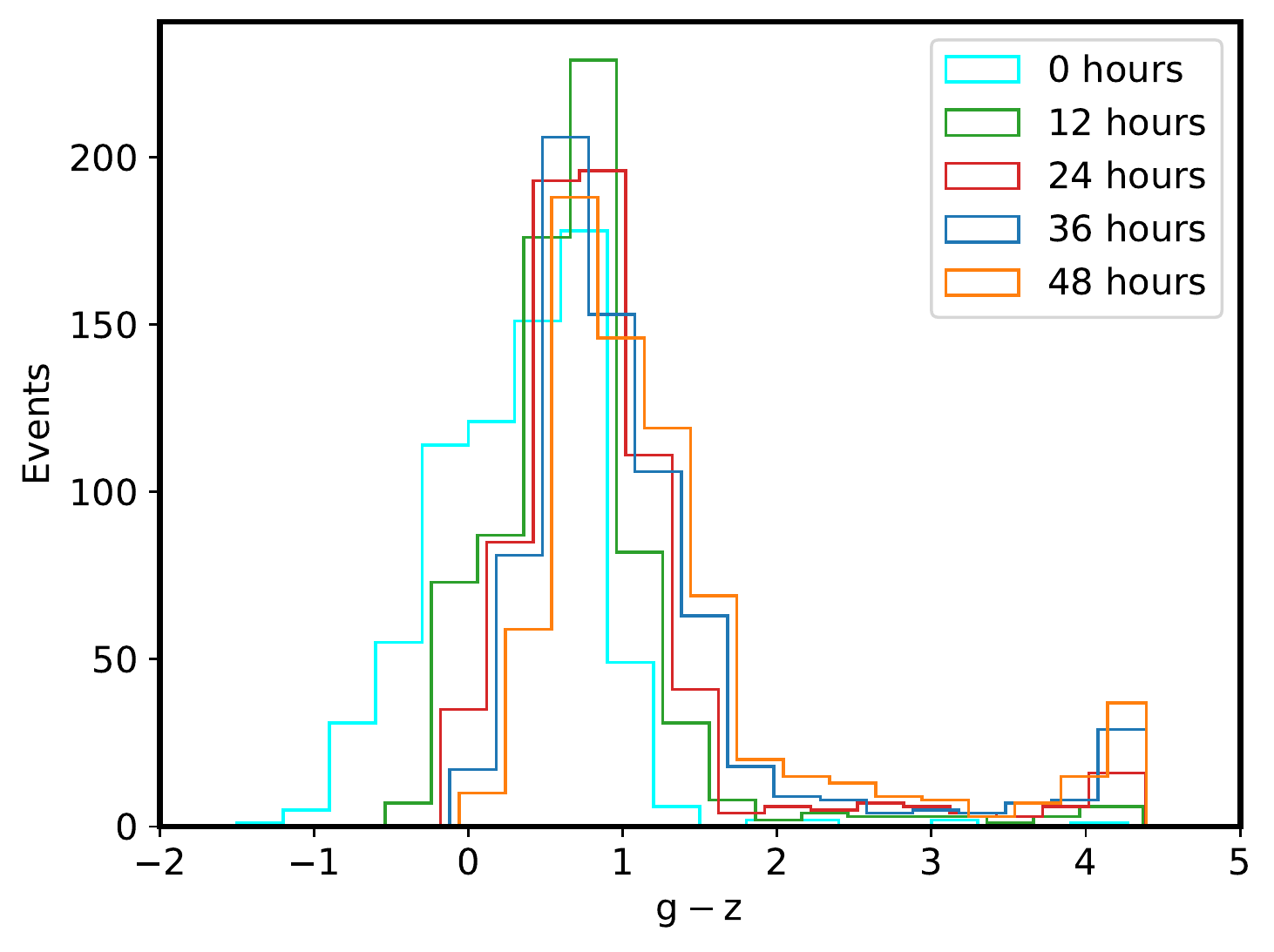}
    \includegraphics[width=\columnwidth]{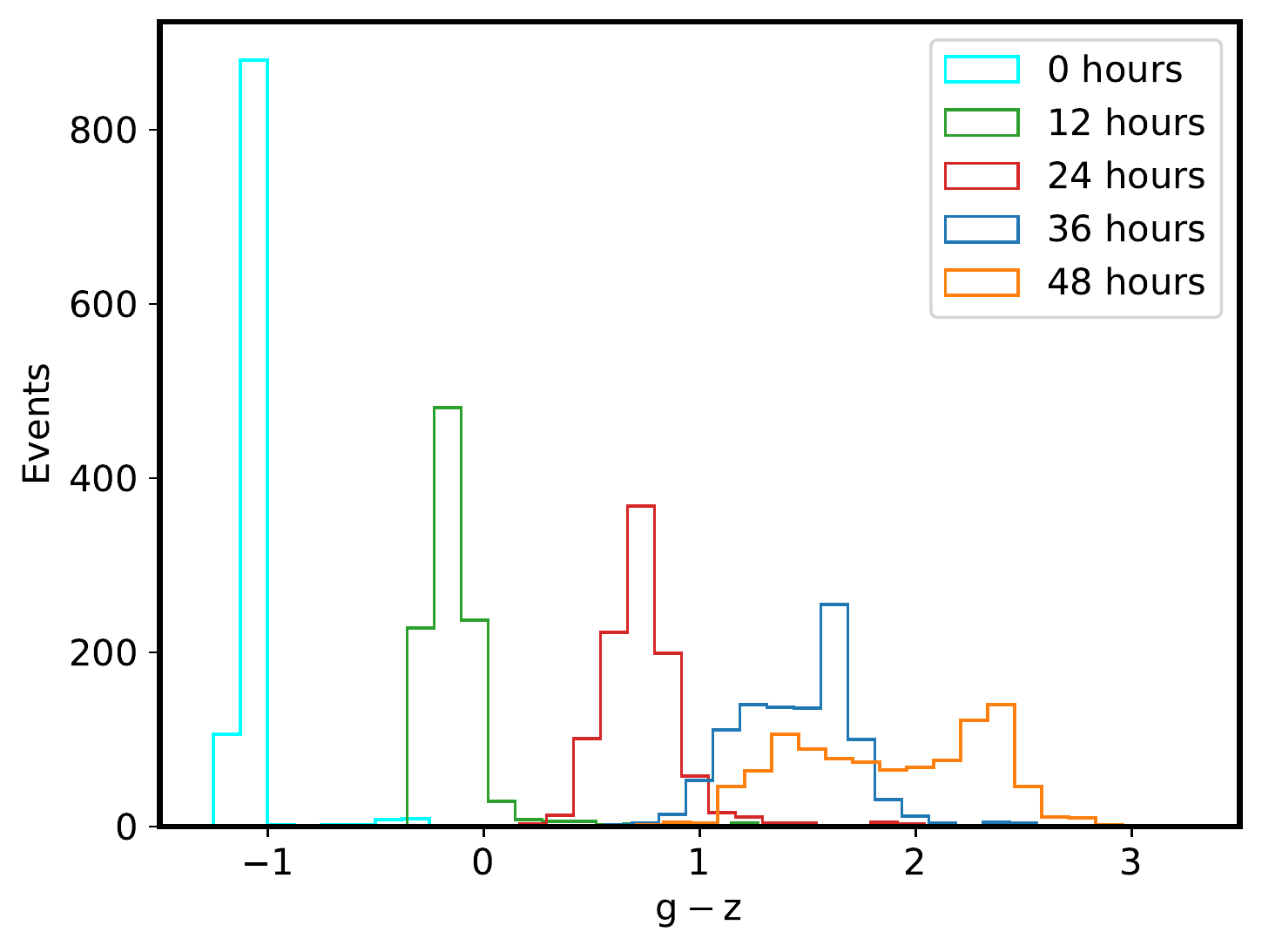}
    \caption{The colour evolution of NSBH kilonovae (left) and BNS kilonovae (right). The histograms show the distribution of g-z colours for the simulated populations at time intervals of 12 hours following the merger. Due to the relative paucity of blue emission, the NSBH model predicts very little colour evolution and a consistently red transient. By contrast, the BNS model predicts rapid blue-to-red evolution over the first two days.}
    \label{fig:colour}
\end{figure*}

In Figure~\ref{fig:colour}, we show the expected colour evolution of the NSBH and BNS kilonovae. Notably, the NSBH kilonovae show little colour evolution, with a consistent g-z colour distribution centred around 1. Conversely, the BNS kilonovae are seen to evolve rapidly in colour over the first two days, becoming comparable to the NSBH kilonovae after $\sim 24$ hours. This is likely a product of the lack of `blue' emission from NSBH mergers, and reinforces the need for early observations to distinguish between the two in cases where the GW signal can't.

\section{Conclusions}\label{sec:conclusions}

We present a new semi-analytic framework capable of predicting NSBH kilonova light curves from input binary properties. The model is integrated into the {\sc mosfit} platform, and can be used for fast generation of libraries of light curves from an input binary population, predicting EM signals accompanying NSBH GW mergers, or performing multi-messenger parameter inference from GW-EM datasets.

We demonstrate that a fiducial NSBH binary with $M_{\rm BH} = 5$\,M$_{\odot}$ and $M_{\rm NS} = 1.4$\,M$_{\odot}$ is broadly consistent with existing candidate kilonova counterparts to cosmological SGRBs with only minor tuning of parameters. However, we also demonstrate that NSBH systems are not capable of producing `blue' emission (likely from lanthanide-poor ejecta) in quantities sufficient to match the light curve of GW170817 unless other processes like shock heating from a GRB jet are included. Simulations \citep[e.g.][]{Fernandez19} suggest that material may be present in polar regions at the time of jet launch, but it is unclear whether there is sufficient mass to result in a signal similar to the one proposed by \citet{Piro18}. Our model indicates that for our assumed prior distributions, less than a third of NSBH mergers within the LIGO range of $\sim 600$\,Mpc will have EM counterparts detectable with Rubin/LSST, even before accounting for survey cadence and line-of-sight extinction.

Our modelling suggests that early ($\lesssim 2$ days) observations of emergent kilonovae will be essential to distinguish BNS and NSBH mergers in cases where GW signals are absent or ambiguous. We also show that NSBH kilonovae may not peak at optical frequencies until up to a week after merger for certain viewing angles. The first discovery of an EM signal from an NSBH merger remains a key objective of GW-EM and transient astronomy. Its identification will serve to validate (or iterate) merger models, as was done for the BNS case following GW170817. Our model provides an early framework for interpreting the emission from such a system, and a platform for further development following observational ratification.

\section*{Acknowledgements}
We are extremely grateful to Rodrigo Fernandez for fruitful discussions that helped shape the model.

We thank the anonymous referee for their constructive comments that improved the quality of the manuscript.

MN and BG acknowledge funding by the European Research Council (ERC) under the European Union’s Horizon 2020 research and innovation programme (grant agreement No.~948381). GPS acknowledges support from The Royal Society, the Leverhulme Trust, and the Science and Technology Facilities Council (grant numbers ST/N021702/1 and ST/S006141/1).

\section*{Data Availability}

The model is available as part of {\sc mosfit} v1.1.9, and can be accessed at \href{https://github.com/guillochon/MOSFiT}{https://github.com/guillochon/MOSFiT}. Installation and general usage instructions are available at \href{http://mosfit.readthedocs.io/}{http://mosfit.readthedocs.io/}.



\bibliographystyle{mnras}
\bibliography{manuscript} 




\appendix



\bsp	
\label{lastpage}
\end{document}